\documentclass[12pt,a4paper]{article}
\usepackage{amsmath,caption,amsfonts,setspace,mathrsfs,mathtools}
\usepackage[top=1.4in, bottom=1.4in, left=1in, right=1in]{geometry}

\usepackage[round]{natbib}
\usepackage{color,soul}
\DeclareCaptionStyle{italic}[justification=centering]{labelfont={bf},textfont={it},labelsep=colon}
\usepackage[T1]{fontenc}
\usepackage[latin9]{inputenc}

\usepackage{graphicx,psfrag,epsf,textcomp}
\usepackage{enumerate, dsfont}
\usepackage{natbib}
\usepackage{url,xcolor} 
\usepackage{booktabs, subfig, bm, paralist,mathpazo,tikz,todonotes,longtable,microtype}

\usepackage[pdftex,colorlinks=true]{hyperref}
\definecolor{darkblue}{rgb}{0,0,.6}
\hypersetup{citecolor=darkblue,linkcolor=darkblue,urlcolor=darkblue}
\newcommand{\argmax}{\operatornamewithlimits{argmax}}

\newcommand{\blind}{0}

\providecommand{\tabularnewline}{\\}
\newcommand\possessivecite[1]{\citeauthor{#1}'s \citeyearpar{#1}}

\graphicspath{{plots/}}

\newsavebox\CBox
\def\textBF#1{\sbox\CBox{#1}\resizebox{\wd\CBox}{\ht\CBox}{\textbf{#1}}}

\begin{document}

\def\spacingset#1{\renewcommand{\baselinestretch}{#1}\small\normalsize} \spacingset{1}

\if0\blind
{
  \title{\bf Implied Volatility Surface Predictability: \mbox{The Case of Commodity Markets}}
  \author{
  Fearghal Kearney\footnote{Corresponding Author: Queen\textquoteright s Management School, Queen\textquoteright s University Belfast, BT9 5EE, UK. Telephone: +44 28 9097 4795. E-mail: f.kearney@qub.ac.uk} \\
  \\
  Han Lin Shang\footnote{Research School of Finance, Actuarial Studies and Statistics, Australian National University, Canberra ACT 2601, Australia. E-mail: hanlin.shang@anu.edu.au} \\
  \\
  Lisa Sheenan\footnote{Queen\textquoteright s Management School, Queen\textquoteright s University Belfast, BT9 5EE, UK. E-mail: l.sheenan@qub.ac.uk}
  \\
 }
  \maketitle
} \fi

\if1\blind
{
	\title{\bf Implied Volatility Surface Predictability: \mbox{The Case of Commodity Markets}}
\maketitle
} \fi

\bigskip
\begin{abstract}
Recent literature seek to forecast implied volatility derived from equity, index, foreign exchange, and interest rate options using latent factor and parametric frameworks. Motivated by increased public attention borne out of the financialization of futures markets in the early 2000s, we investigate if these extant models can uncover predictable patterns in the implied volatility surfaces of the most actively traded commodity options between 2006 and 2016. Adopting a rolling out-of-sample forecasting framework that addresses the common multiple comparisons problem, we establish that, for energy and precious metals options, explicitly modeling the term structure of implied volatility using the Nelson-Siegel factors produces the most accurate forecasts.
\\

\noindent Keywords: Implied volatility surfaces; Options markets; Forecasting; Commodity finance {\scriptsize{}\medskip{}}{\scriptsize \par}
\noindent JEL Classifications: G10; G15; G17{\scriptsize{}\bigskip{}}{\scriptsize \par}
\end{abstract}

\newpage

\def\spacingset#1{\renewcommand{\baselinestretch}{#1}\small\normalsize} \spacingset{1}
\spacingset{1.45}

\newpage

\section{Introduction}\label{sec:Introduction}

When using the \cite{black1973pricing} model to price options, the only variable not known with certainty is volatility. The estimated future volatility backed out of these option prices is referred to as implied volatility (IV). This can be plotted against both moneyness and time-to-maturity to produce an implied volatility surface (IVS). While \cite{black1973pricing} assume that the IVS is flat, this is not observed empirically, as option contracts of varying maturity and moneyness levels tend to be priced according to different levels of IV. The shape of these IVSs evolves, and so accurate modeling of their inter-temporal dynamics is required \citep{deuskar2008economic, konstantinidi2008can, BH09}. \cite{BH09} state that producing reliable forecasts for the evolution of IV is essential to indicate prevailing market conditions as well as to facilitate efficient option portfolio risk management. Research to date focuses on modeling and forecasting equity, index, interest rate, and foreign exchange options, while uncovering predictable dynamics in commodity IV has yet to be explored in the literature. We aim to address this research gap.

\cite{irwin2011index} report that flows into commodity investments increased from \$15 billion in 2003 to \$250 billion in 2009. \cite{fattouh2014causes} cite this \textit{financialization} of futures markets as the starting point for increased public interest in commodities.\footnote{Financialization is the term commonly used for the phenomenon in the early 2000s whereby large inflows into commodity investments occurred. \cite{DH07} attribute this to institutional investors who were historically not engaged in commodity investing on such a large scale.} \cite{adams2015financialization} conclude that the entrance of these new types of investors significantly influenced the behavior of commodities in financial markets and how commodities linked to other assets. More specifically, as commodities became a significant part of financial investors\textquoteright{} portfolios, they were treated as a new category within the universe of stocks, becoming part of a more general equity style. A similar view is put forward by \cite{cheng2014financialization}, who postulate that commodity futures are treated as an asset class just like stocks and bonds.\footnote{This recent shift in how commodities are considered represents a stark departure from the description of commodity returns as being weakly correlated with the stock market \citep{bessembinder1992time, gorton2006facts}.} Given this interconnectedness between commodities and other asset classes, we investigate how readily extendable to commodity markets the methods popular for modeling and forecasting the IVSs of other assets are. Specifically, we determine how accurately the existing frameworks characterize the shape of the commodity IVS and how well they uncover predictable patterns and dynamics unique to these underlying assets.

Related literature highlights a multitude of reasons why forecasting IV is important for asset pricing and risk management. As well as being a transformation of the option price, and a key parameter in many asset pricing formulae \citep{Giot03}, IV is of interest because of its use in forecasting realized volatility \citep[see][]{corrado2006estimating, taylor2010information, muzzioli2010option, garvey2012realised}. \cite{Giot03} also demonstrate the high informational content of IV by successfully adopting it in a value-at-risk modeling framework for agricultural futures. \cite{bernales2014can} posit that understanding the dynamics of the IVS can inform traders regarding the design of speculative or hedging strategies. Knowledge of the dynamic process of the IVS is also relevant for investment decisions in other markets because options are commonly used to obtain forward-looking market information. Such forward-looking analyses rely on the assumption that option prices reveal agents\textquoteright{} expectations about prospective economic scenarios, where investor forecast horizons correspond to the expiry dates of traded options contracts.

A variety of frameworks has been proposed to model and forecast IV. We split them into four broad classes: general equilibrium, principal component (PC), machine learning, and parametric modeling. Turning first to general equilibrium models \cite{david2000option, guidolin2003option, garcia2003empirical, hibbert2008behavioral} and \cite{bernales2015learning} develop rational asset pricing models that theoretically calibrate to observed asymmetric IVS shape and its evolution over time. The underlying idea is that investors\textquoteright{} uncertainty about economic fundamentals (e.g., dividends) affects stochastic volatility and leverage behind the IVS. This uncertainty is said to evolve. However, as outlined by \cite{BH09}, these general equilibrium models suffer from tractability issues and are not appropriate for risk management or forecasting purposes.

An alternative approach is to use a purely data-driven method that describes the empirical shape of the IVS through the use of unobservable latent statistical factors, as in \cite{SHC00, alexander2001market,chalamandaris2010predictable}. \cite{chalamandaris2010predictable} exploit PC factors further by employing them to produce out-of-sample forecasts of the IVS. However, these forecasts are produced using a regular-sized daily IVS dataset with precisely the same moneyness and maturity observations available on each observation date. Such an approach is not possible using our trade-level commodity dataset in which contracts with the same moneyness and maturity are not traded every day.

A third related methodology is that of machine learning and non-parametric techniques to model the IVSs, as in \cite{CF02, FHV03, FHM07, AC10, FH15}. The primary intuition behind these models is that, instead of estimating a highly parameterized model, it is better to use non- or semi-parametric techniques with the amount of complexity controlled to avoid overfitting. However, of the studies cited above, only \cite{AC10} produce out-of-sample forecasts of the IVS in line with the aim of our paper. They utilize regression trees informed by a cross-validation strategy to produce forecasts of the surface, an approach that we also follow in our out-of-sample analysis.

A fourth popular methodology involves the use of deterministic parametric specifications based on the cross-section of options available at any one point in time. The specifications link the maturity and moneyness of option contracts to the shape of the IVS, with \cite{dumas1998implied} and \cite{pena1999we} the most popular examples. \cite{goncalves2006predictable} use a two-stage framework to forecast \possessivecite{dumas1998implied} cross-sectional maturity and moneyness coefficients. Further, \cite{chalamandaris2011important} extend their work by explicitly modeling the IVS term structure component by incorporating \possessivecite{nelson1987parsimonious} factors.

These forecasting frameworks have been applied across a wide number of markets including equities, foreign exchange, and interest rates. For example, \cite{goncalves2006predictable} describe the daily IVS of S\&P500 Index options using \possessivecite{dumas1998implied} parametric specifications. They interpret the daily estimated parameter coefficients as proxies for factors that drive the evolution of the IVS and empirically demonstrate that modeling the time series dynamics can identify predictable IVS components. \cite{BH09} also focus on S\&P500 Index options, instead of implementing parameter estimation using a Kalman filter. However, they focus solely on information from an isolated IV smile and do not incorporate information from the full surface as we seek to do. \cite{FHM07}, also model equity index options, namely the German stock index (DAX) using their semi-parametric factor model. After selecting an appropriate model size and bandwidth, it is proposed that the estimated factors series could follow a vector autoregressive (VAR) model. However, no direct out-of-sample forecasts of the IV surface are produced. \cite{chalamandaris2010predictable, chalamandaris2011important, Chalamandaris2014} examine foreign exchange markets. For instance, \cite{Chalamandaris2014} use an extensive time series of over-the-counter (OTC) options for eight currencies versus the Euro. They find that in medium- to long-term forecasts the shape and dynamics of the IVS can be forecast successfully using the aforementioned latent factor and parametric models. Finally, \cite{deuskar2008economic} analyze the interest rate options market, finding that the shape of the IV smile for Euro interest rate caps and floors is affected dynamically by the yield curve and by future uncertainty in interest rate markets.\footnote{\noindent \cite{deuskar2008economic} also highlight that conclusions from equity markets do not necessarily transfer to interest rate options as they are predominantly traded OTC by asking side institutional investors. However, such constraints do not apply to commodity options as futures contracts are the most active avenue via which investors take positions in the market.}\footnote{\noindent In a related study of IV indices by \cite{badshah2013contemporaneous}, casual contemporaneous bidirectional spill-over between gold and exchange rates are uncovered. This is a tentative indication that frameworks adopted to model the dynamics of foreign exchange markets may also apply to Precious Metal commodities.}

There is a growing body of evidence for predictable IVS movements. Predicting the entire IVS, and thus all future option prices, would directly contradict the efficient market hypothesis. However, uncovering predictability in isolated segments of a commodity IVS may signal the existence of important pockets of market inefficiency. \cite{Chalamandaris2014} conclude that non-uniform trading across the IVS may lead to some segments exhibiting more predictability than others as they adjust to information at different speeds. Further, \cite{chalamandaris2011important} argue that over longer horizons, the shape of the IVS term structure contains predictive information, whereby forward IVs may be utilized to forecast future implied volatilities of nearer expiries. If the efficient market hypothesis is imposed, this IVS predictability may be traced back to either micro-structural imperfections or unobservable and hard-to-estimate time-varying risk premium.

As in \cite{Chalamandaris2014} we choose popular models from the literature: the two-step approach from \cite{goncalves2006predictable}, and the \cite{diebold2006forecasting} based approach of \cite{chalamandaris2011important}. We apply these frameworks along with a regression tree benchmark model from \cite{AC10}, to a representative sample of the most actively traded commodity options contracts: Cocoa, Corn, Cotton, Soybean, Soybean Oil, Sugar, Wheat, Crude Oil, Heating Oil, Natural Gas, Gold, and Silver futures. We seek to model the shape of the observed IVS in the context of this distinct market. We also employ a rolling out-of-sample framework to identify statistically significant forecasting performance. Through an assessment of these models in a multiple comparisons setting, we formally establish the set of superior approaches. We find that in a cross-model comparison of parametric and machine learning approaches no one class of model systematically outperforms. The most striking observation is that the \cite{goncalves2006predictable} model performs poorly overall, indicating that using smile shape factors only and a linear approximation of the term structure dimension is not sufficient to characterize the dynamics observed in commodity options. However, the results for different classes of commodities are more illuminating, in that for Energy and Precious Metals, the \cite{chalamandaris2011important} framework produces the most accurate forecasts. Conversely the theory-free machine learning regression tree approach exhibits the most promising results for Agricultural options, in particular those models with flatter underlying futures convenience yields.

The paper proceeds as follows: Section~\ref{sec:data set} presents the data used in the study. Section~\ref{sec:Methodology} outlines the models used to characterize and forecast the IVS. Section~\ref{sec:Empirical Findings} provides an empirical assessment of out-of-sample testing. Finally, Section~\ref{sec:Conclusions} concludes and suggests avenues for potential future research.

\section{The Data}\label{sec:data set}

The dataset is obtained from ivolatility.com and contains IV quotes for a representative sample of the most actively traded commodity options, covering Agricultural, Energy and Precious Metals. The primary data are from ACTIV Financial Systems, Inc., which is an Options Price Reporting Authority-approved vendor, with data incorporating traded-level information including volume. Specifically, it includes options data for Cocoa, Corn, Cotton, Soybean, Soybean Oil, Sugar, Wheat, Crude Oil, Heating Oil, Natural Gas, Gold, Copper, and Silver futures. The commodities can be grouped into three broad classes: Agricultural (Cocoa, Corn, Cotton, Soybean, Soybean Oil, Sugar, and Wheat); Energy (Crude Oil, Heating Oil, and Natural Gas); and Precious Metals (Copper, Gold, and Silver). The sampling frequency is daily and the sample period runs from January 2006 to December 2016. We consider both call and put options but only include contracts that have at least one transaction on a given day, that is volume $>0$. As one would expect we do not have the same number of traded contracts on every observation date. To enhance data quality and to ensure that all segments of the surface are actively traded, we first consider only options in the 90-110\% moneyness range and with maturities of between one month and two years. Second, to eliminate any remaining bias arising from thinly traded expiration dates, we concentrate on the most liquid maturities by imposing a filtering rule. This rule requires at least 15,000 quotes within each maturity group of 1-6 months, 6-12 months, 12-18 months, and 18-24 months, for that maturity group to be included in our dataset. This filtering criterion ensures that we have at least five IV quotes for each day for each maturity group. However, the vast majority of days across the commodities in our sample contain many multiples of this. Unfortunately, after filtering the data, Copper options have an insufficient number of contracts traded and are therefore dropped from our dataset. The results of this filtering exercise, as well as the descriptive statistics for the dataset, are reported in  Table~\ref{tab:Data-Description}. Mean IV levels lie in the 19.03-36.65\% range with distributions displaying a predominantly positive skew, in line with the commodity options literature \citep{mandelbrot1963new, fama1965behavior}. 

Further, we include an indicator of the term structure of the underlying futures curves for each of our commodities through the inclusion of an average convenience yield slope. The convenience yield slope provides an indication of both the cost of carrying, in our case most often storage and insurance, and also the market expectation of future price trajectory. We can see how this varies across commodities as strong futures curves dynamics are observed for Soybeans and Wheat, with average slopes of -5.21 and 3.71, respectively.

\begin{table}[!htbp]
\caption{Data Description}\label{tab:Data-Description}
\centering
\tabcolsep 0.025in
\begin{tabular}{@{}llcccccccc@{}}
\toprule
{\footnotesize{}Ticker} & {\footnotesize{}Commodity} & {\footnotesize{}Mean} & {\footnotesize{}Median} & {\footnotesize{}Std. Dev.} & {\footnotesize{}1st Quart.} & {\footnotesize{}3rd Quart.} & {\footnotesize{}Skewness} &  {\footnotesize{}Longest Mat.} & {\footnotesize{}Avg. Convenience}\tabularnewline
& {\footnotesize{}(Exchange)} &(\%)&(\%)&(\%)&(\%)&(\%)& & & {\footnotesize{}Yield Slope} \\
\midrule 
{\footnotesize{}CC} & {\footnotesize{}Cocoa (ICE)} & {\footnotesize{}26.76} & {\footnotesize{}24.38} & {\footnotesize{}6.79} & {\footnotesize{}21.66} & {\footnotesize{}31.74} & {\footnotesize{}80.43} & {\footnotesize{}6 Months} & {\footnotesize{}1.08}\tabularnewline
{\footnotesize{}C} & {\footnotesize{}Corn (CBOT)} & {\footnotesize{}29.35} & {\footnotesize{}28.08} & {\footnotesize{}7.12} & {\footnotesize{}23.89} & {\footnotesize{}34.22} & {\footnotesize{}57.16} & {\footnotesize{}12 Months} & {\footnotesize{}0.31}\tabularnewline
{\footnotesize{}CT} & {\footnotesize{}Cotton (ICE)} & {\footnotesize{}27.24} & {\footnotesize{}24.15} & {\footnotesize{}9.15} & {\footnotesize{}20.93} & {\footnotesize{}30.00} & {\footnotesize{}166.07} & {\footnotesize{}12 Months} & {\footnotesize{}-0.55}\tabularnewline
{\footnotesize{}S} & {\footnotesize{}Soybeans (CBOT)} & {\footnotesize{}23.56} & {\footnotesize{}21.87} & {\footnotesize{}6.63} & {\footnotesize{}19.33} & {\footnotesize{}25.75} & {\footnotesize{}146.45} & {\footnotesize{}12 Months} & {\footnotesize{}-5.21}\tabularnewline
{\footnotesize{}BO} & {\footnotesize{}Soybean Oil (CBOT)} & {\footnotesize{}22.30} & {\footnotesize{}21.05} & {\footnotesize{}5.06} & {\footnotesize{}18.98} & {\footnotesize{}23.72} & {\footnotesize{}164.42}  & {\footnotesize{}6 Months} & {\footnotesize{}0.15}\tabularnewline
{\footnotesize{}SB} & {\footnotesize{}Sugar (ICE)} & {\footnotesize{}32.36} & {\footnotesize{}30.39} & {\footnotesize{}10.57} & {\footnotesize{}24.52} & {\footnotesize{}39.56} & {\footnotesize{}58.35}  & {\footnotesize{}6 Months} & {\footnotesize{}-0.05}\tabularnewline
{\footnotesize{}W} & {\footnotesize{}Wheat (CBOT)} & {\footnotesize{}30.59} & {\footnotesize{}28.87} & {\footnotesize{}7.54} & {\footnotesize{}24.87} & {\footnotesize{}35.80} & {\footnotesize{}79.51}  & {\footnotesize{}12 Months} & {\footnotesize{}3.71}\tabularnewline
{\footnotesize{}CL} & {\footnotesize{}Crude Oil (NYMEX)} & {\footnotesize{}31.48} & {\footnotesize{}30.24} & {\footnotesize{}10.70} & {\footnotesize{}24.99} & {\footnotesize{}36.96} & {\footnotesize{}121.20}  & {\footnotesize{}24 Months} & {\footnotesize{}0.12}\tabularnewline
{\footnotesize{}HO} & {\footnotesize{}Heating Oil (NYMEX)} & {\footnotesize{}27.71} & {\footnotesize{}28.09} & {\footnotesize{}8.20} & {\footnotesize{}20.61} & {\footnotesize{}33.12} & {\footnotesize{}37.02} & {\footnotesize{}6 Months} & {\footnotesize{}0.01}\tabularnewline
{\footnotesize{}NG} & {\footnotesize{}Natural Gas (NYMEX)} & {\footnotesize{}36.65} & {\footnotesize{}34.53} & {\footnotesize{}9.72} & {\footnotesize{}29.77} & {\footnotesize{}41.67} & {\footnotesize{}123.95}  & {\footnotesize{}24 Months} & {\footnotesize{}0.06}\tabularnewline
{\footnotesize{}GC} & {\footnotesize{}Gold (COMEX)} & {\footnotesize{}19.03} & {\footnotesize{}17.81} & {\footnotesize{}5.27} & {\footnotesize{}15.61} & {\footnotesize{}21.03} & {\footnotesize{}199.60}  & {\footnotesize{}12 Months} & {\footnotesize{}0.99}\tabularnewline
{\footnotesize{}SI} & {\footnotesize{}Silver (COMEX)} & {\footnotesize{}31.82} & {\footnotesize{}30.33} & {\footnotesize{}8.04} & {\footnotesize{}26.21} & {\footnotesize{}35.71} & {\footnotesize{}124.25}  & {\footnotesize{}12 Months} & {\footnotesize{}0.02}\tabularnewline
\bottomrule 
\end{tabular}
{\footnotesize For each of the 12 commodity options in our sample, this table reports descriptive statistics for the implied volatilities, essentially the market's expectation of future volatility between now and the option expiry date. It reports the ticker symbol used, the exchange the commodity trades on, the longest option maturity considered after filtering criteria is applied, and the average convenience yield slope of the underlying futures contract. The sample period is January 2006-December 2016.}
\end{table}

\begin{figure}[!htbp]
\caption{Average Implied Volatility Surfaces}\label{fig:Average-Implied-Volatility}
\begin{centering}
\includegraphics[width=8.3cm]{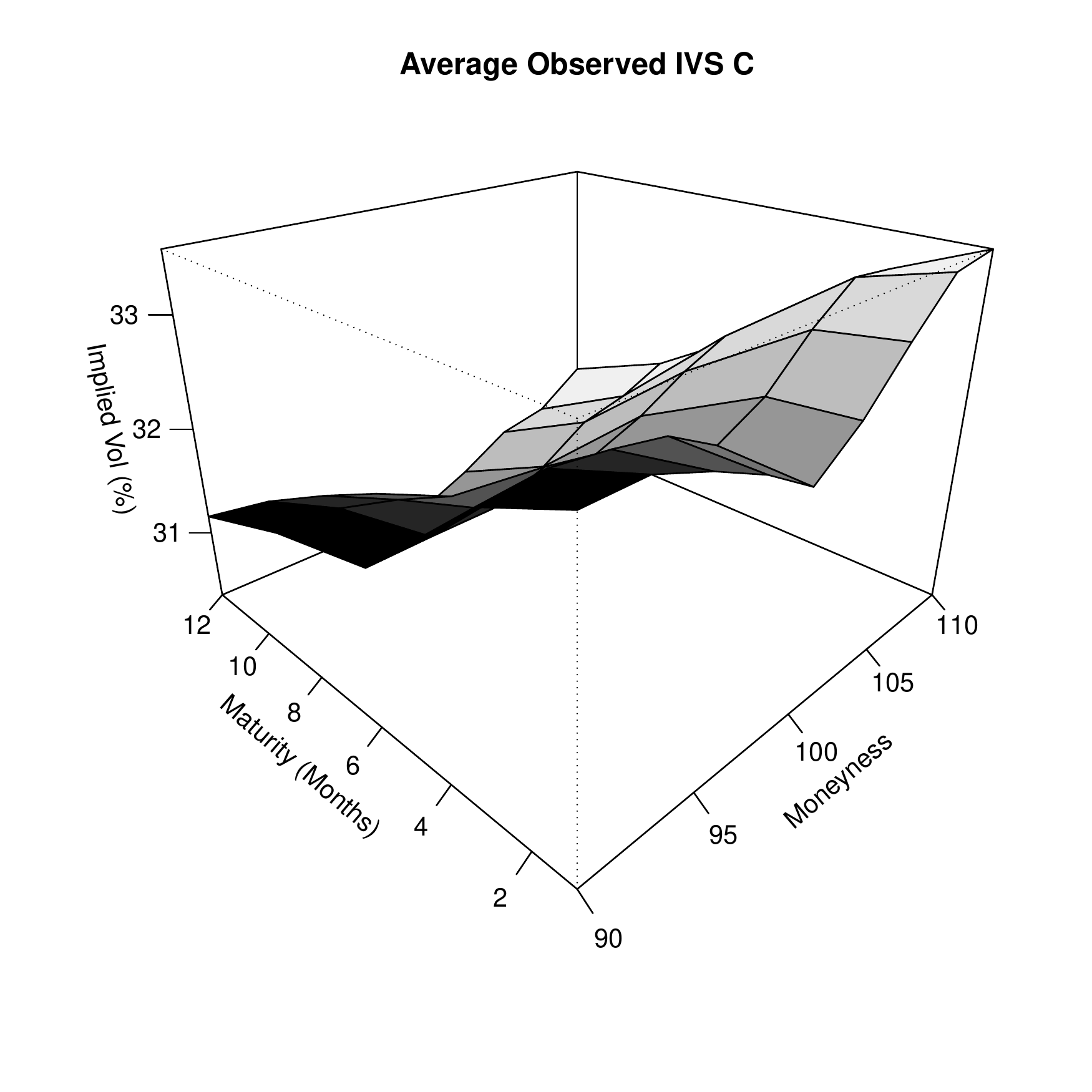}\quad
\includegraphics[width=8.3cm]{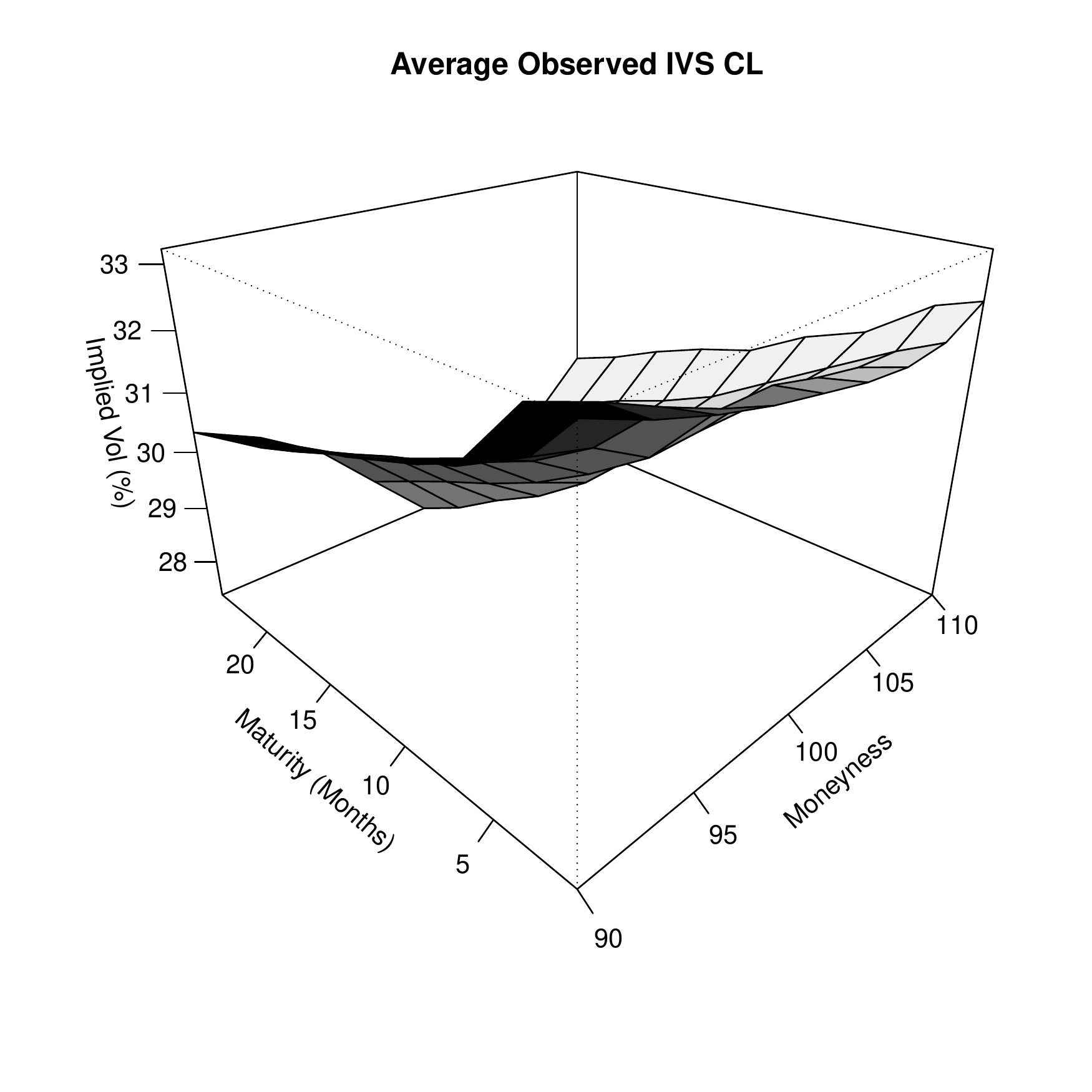}
\par\end{centering}
\begin{centering}
\includegraphics[width=8.3cm]{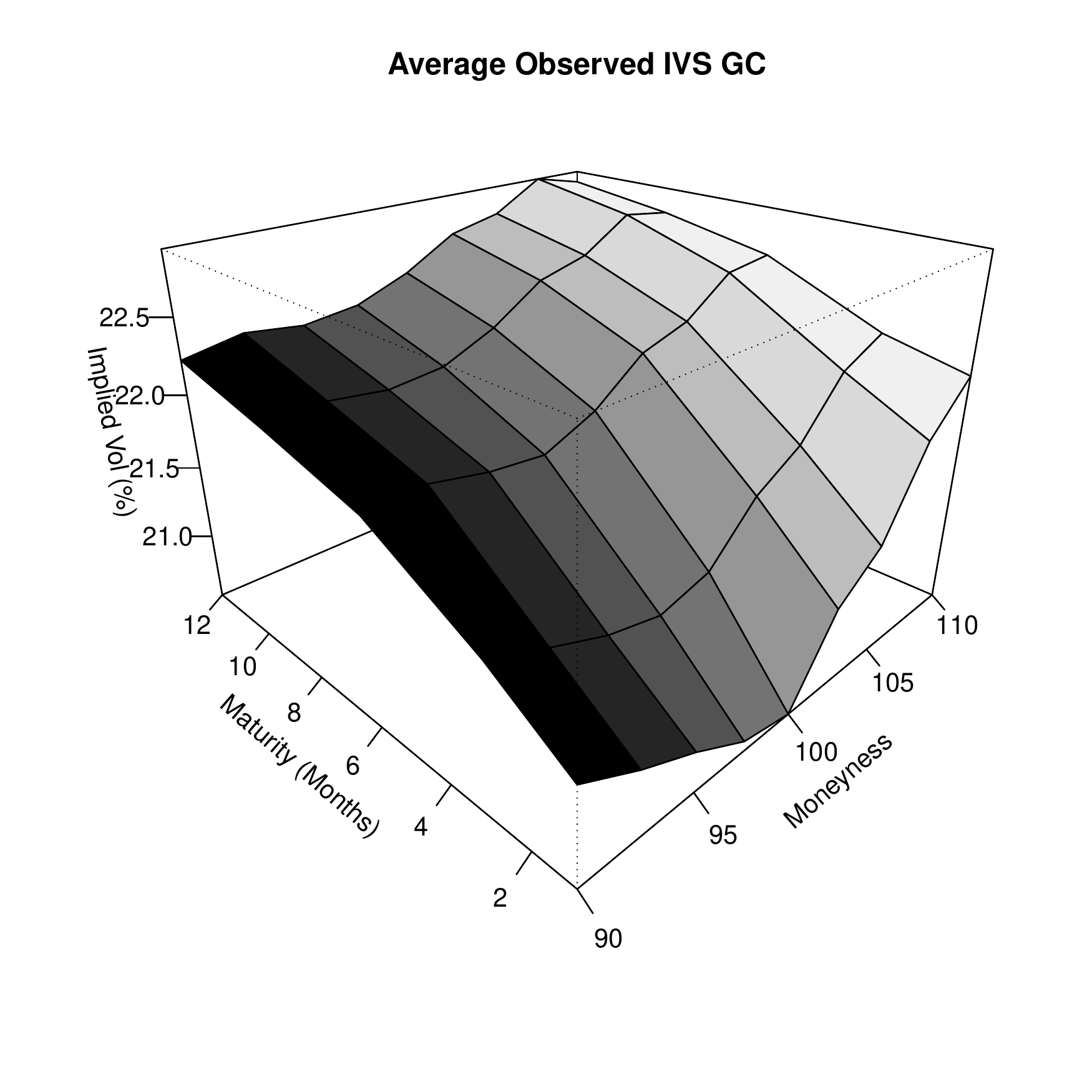}\quad
\includegraphics[width=8.3cm]{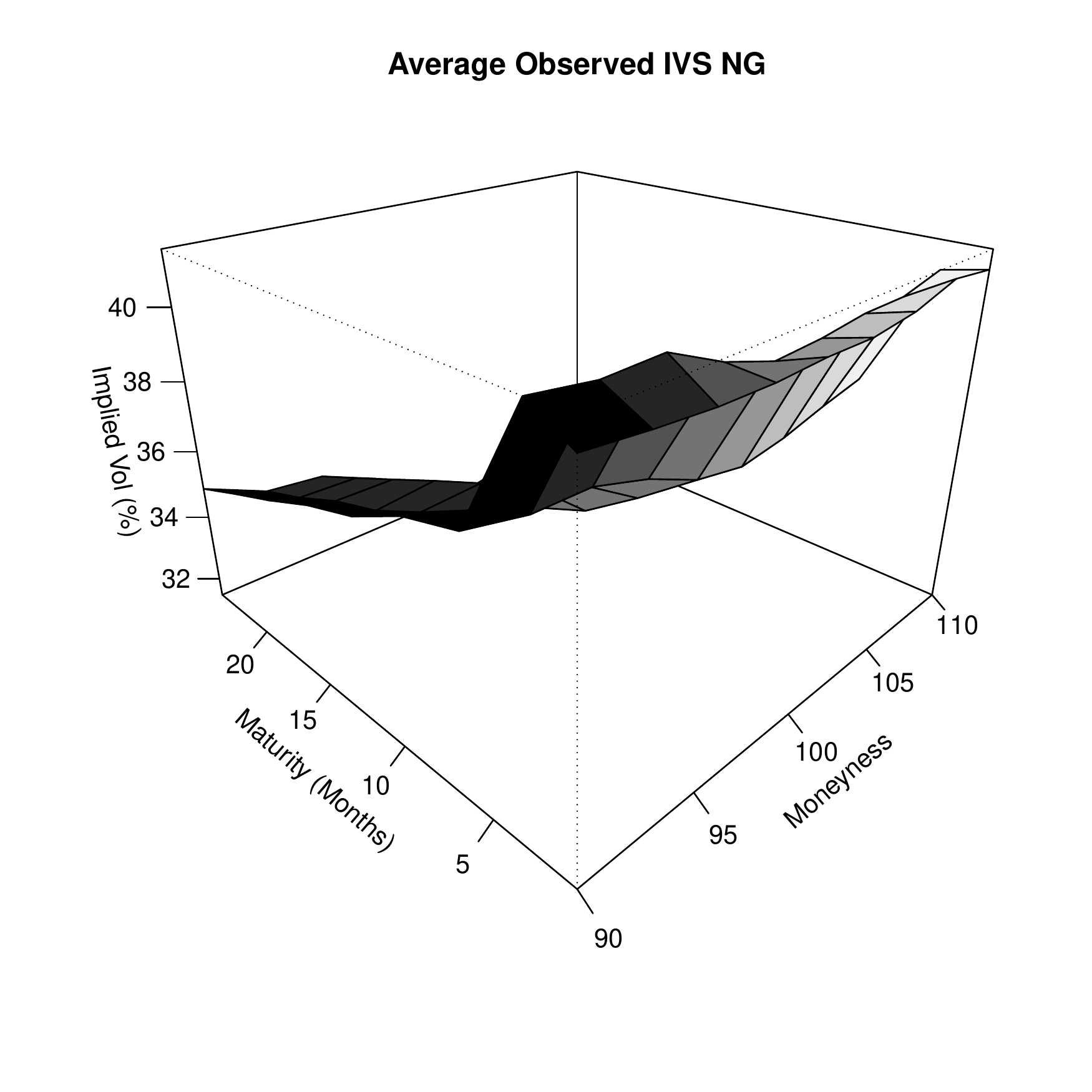}
\par\end{centering}
\begin{centering}
\bigskip{}
\par\end{centering}
\centering{}{\footnotesize{}This figure presents the average IVS plotted for each class of commodity over the January 2006-December 2014 in-sample period. In clockwise order: Corn, Crude Oil, Gold, and Natural Gas. The IVS is a three-dimensional plot where the $x$-axis is the time-to-maturity, the $z$-axis is the moneyness, and the $y$-axis is the implied volatility derived from the option contracts.}{\footnotesize \par}
\end{figure}

Figure~\ref{fig:Average-Implied-Volatility} presents the average IVS for four heavily traded options (Corn, Gold, Crude Oil, and Natural Gas) over the in-sample 2006-2014 period. The average IVS for Corn is presented in the upper left quadrant. We observe a positive skew in the moneyness dimension, whereby out-of-the-money options are more expensive than their in-the-money counterparts, implying fear of future supply-side disruption, as outlined by \cite{askari2008oil}, \cite{liu2011stochastic} and \cite{kearney2015north}. There is also evidence of a premium for short-term maturity options. This suggests that the market anticipates the impact of such supply-side disruptions to even out over a longer time-frame.

\begin{figure}[!htbp]
\caption{Gold Implied Volatility Evolution 2010}\label{fig:Evolution-Implied-Volatility}
\begin{centering}
\includegraphics[width=8.5cm]{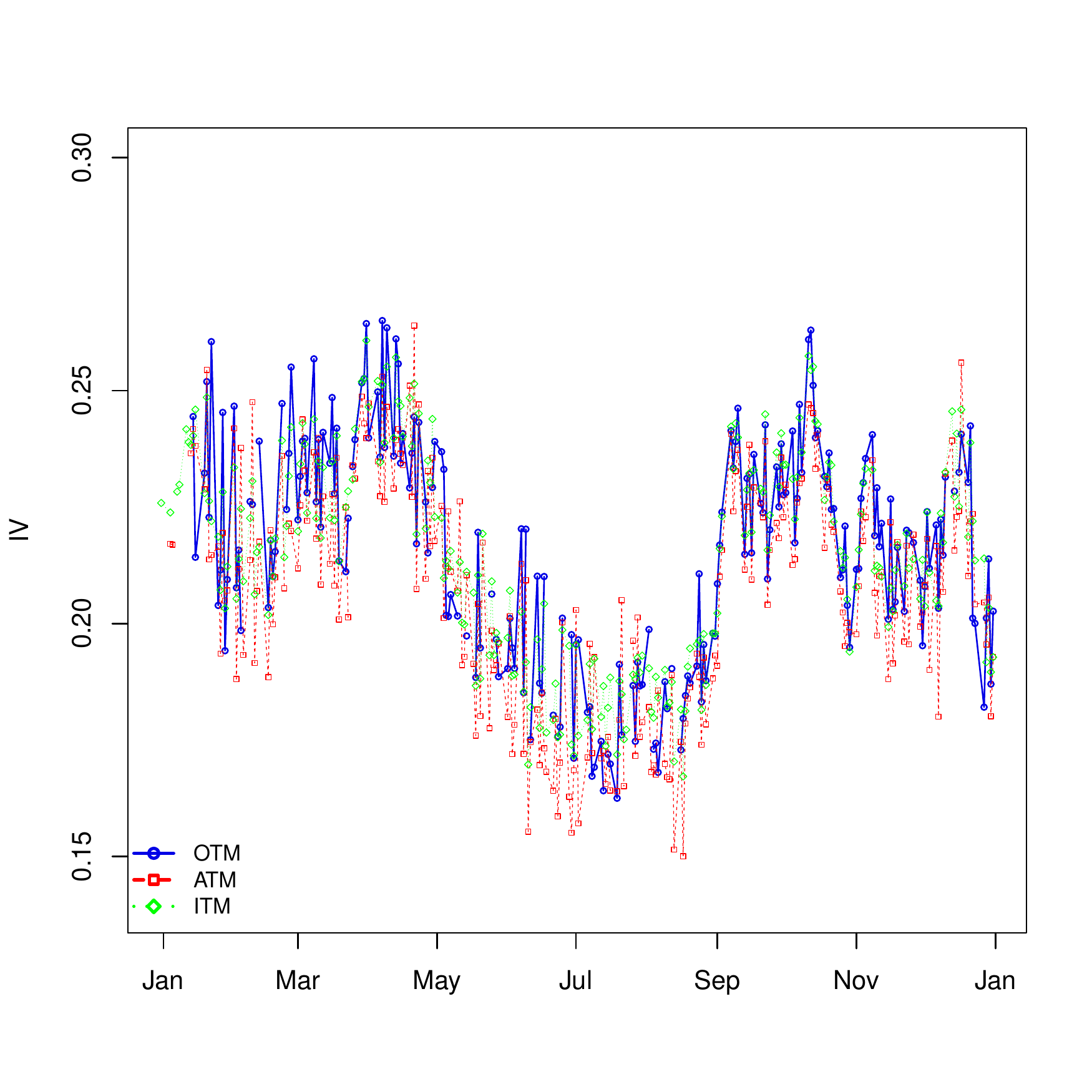}\qquad
\includegraphics[width=8.5cm]{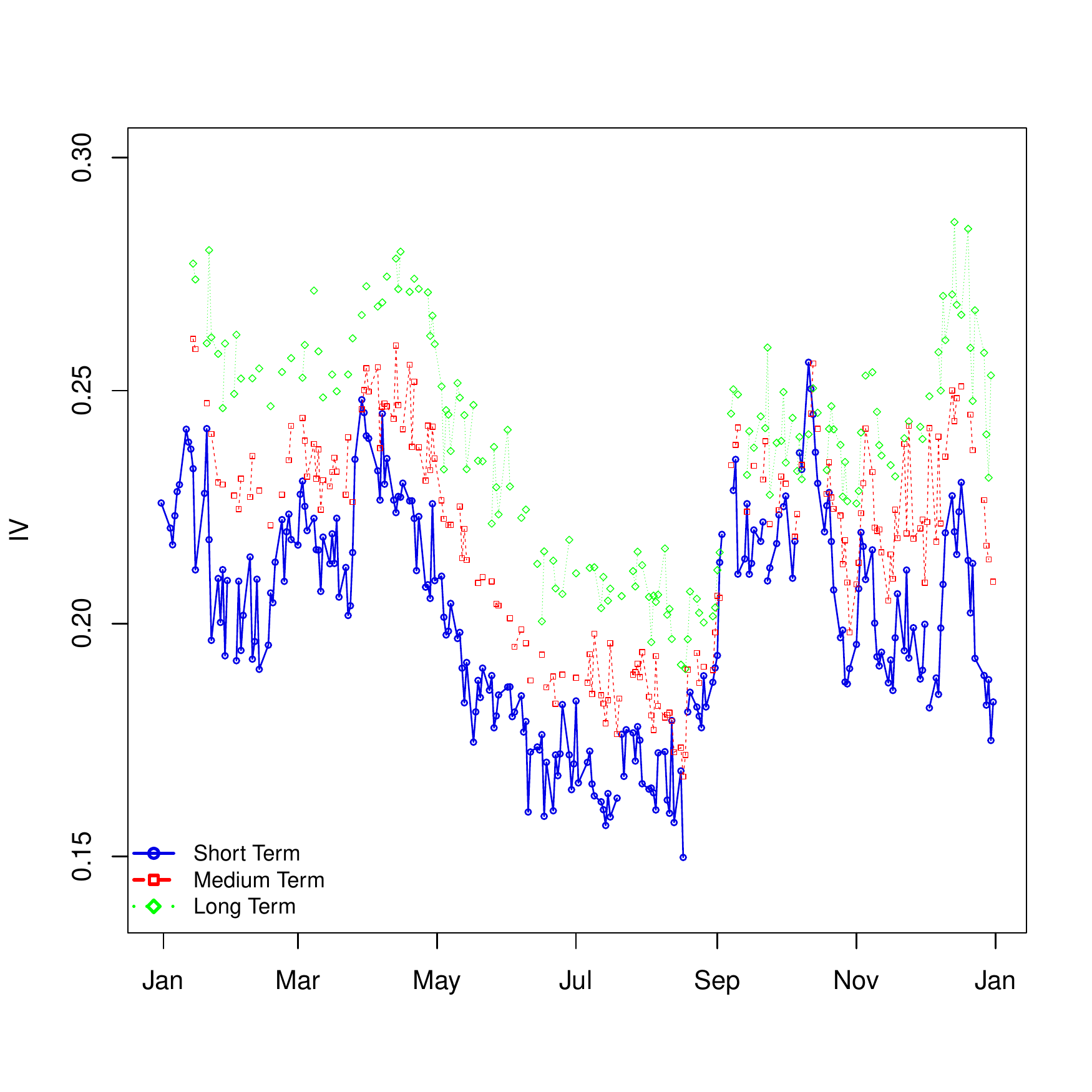}
\par\end{centering}
\begin{centering}
\bigskip{}
\par\end{centering}
\centering{}{\footnotesize{}This figure presents the evolution of the average IV for gold commodity options over the year 2010. It is split by both moneyness and maturity: Out-of-the-money (OTM) (<95\%), At-the-money (ATM) (97.5\% to 102.5\%) and In-the-money (ITM) (>105\%); Short- (one to three months), Medium-(three to six months) and Long-term ($>$ six months) contracts.}{\footnotesize \par}
\end{figure}

Precious Metals are often viewed as a separate asset in their own right, given that they display unique attributes in comparison to other commodities. Such bespoke properties can be observed here when focusing on the average Gold IVS. Despite being predominantly positively skewed the IV smile displays a complex structure. The plot also reveals the non-linear shape of the term structure component. As the option maturities increase up to 12 months, the associated IVs also increase. This is in contrast to the dynamics of the other commodity classes. The dynamic corresponds to a relatively high average underlying futures convenience yield curve slope, at 0.99. This yield curve slope indicates that futures on gold contracts further in the future are more expensive than those futures that are closer to maturity.

Finally, we return to the Energy class, with the average IVS for Natural Gas and Oil options presented in the top right and lower right quadrants of Figure~\ref{fig:Average-Implied-Volatility}. Interestingly, despite being Energy commodities where one would expect fear of supply-side disruptions to dominate, in line with \cite{doran2008computing}, the primary driver during our in-sample period is fear of demand-side shock. This is inferred by the slight negative moneyness skew observed and could be due to the weak economic environment caused by the global financial crisis and its aftermath depressing demand for fossil fuel use. Furthermore, there is visually significant curvature across the maturities in line with \cite{doran2008computing}, who highlight the presence of such curvature in Energy markets. In the next section, we outline models to capture the interesting IVS dynamics presented here.

While useful for understanding the predominant dynamics observed within an individual surface, the above figures do not show us how the surfaces evolve. In contrast, Figure \ref{fig:Evolution-Implied-Volatility} plots the evolution of different segments of the IVS over time. It is presented for Gold option contracts broken down first by moneyness and second by maturity. The groups are as follows: out-of-the-money (OTM) (<95\%), at-the-money (ATM) (97.5-102.5\%), and in-the-money (ITM) (>105\%), Short- (1 to 3 months), Medium- (3 to 6 months), and Long-term (<6 months) contracts. We illustrate the evolution of IV for Gold commodities in the year 2010 in isolation to ensure an appropriate graphical resolution for the dynamics.\footnote{Plots for other time periods and commodity options are available from the authors upon request.} From these plots, we can see that the average level of IV evolves greatly over a single-year period and that the relative relationships between the different moneyness and maturity groups do not stay constant. For instance, the January-March period at the beginning of the plot shows a noticeably defined order in terms of long-term maturities being associated with high IV, relative to shorter-term maturities. However, this relationship does not hold in the later October-November period. The complexity of these dynamics, even on an aggregated basis such as that presented here, motivates the need to explicitly model the evolution of the full IVS.

\section{Methodology}\label{sec:Methodology}

\subsection{Modeling and Forecasting the Surface}

We now outline three general frameworks to describe and forecast the IVS. First, \cite{goncalves2006predictable} use a parametric approach, in that they model daily IV surfaces using \cite{dumas1998implied} specifications based on moneyness and time-to-maturity. Second, \cite{chalamandaris2011important} extend \cite{dumas1998implied} by concluding that the linear approximation of maturity is insufficient. They instead propose the use of Nelson-Siegel term structure factors in the spirit of \cite{diebold2006forecasting} to produce fitted IV surfaces. Forecasts from both these parametric models are produced in a two-step framework, as outlined in Section \ref{sec:Out of Sample Forecasting}, by assuming that the estimated coefficients evolve according to a specified time series process. This allows us to construct a forecasted IVS using these predicted coefficients. Finally, we employ the regression tree benchmark model from \cite{AC10}. We now present each specification in more detail.

\cite{goncalves2006predictable} propose a parametric specification based on moneyness and time-to-maturity to characterize the IVS. More precisely, they employ the following model, based on \cite{dumas1998implied}:
\[
\sigma_{i,t}=\alpha_{0,t}+\alpha_{1,t}\Delta_{i}+\alpha_{2,t}\Delta_{i}^{2}+\alpha_{3,t}\tau_{i}+\alpha_{4,t}\left(\Delta_{i}\times\tau_{i}\right)+\epsilon_{i,t},
\]
where $\Delta_{i}$ is the option moneyness, $\tau_{i}$ is the option time-to-maturity, and $\epsilon_{i,t}$ is a random error term. The estimates, $\widehat{\alpha}_{t}=\left[\widehat{\alpha}_{0,t},\widehat{\alpha}_{1,t},\ldots,\widehat{\alpha}_{4,t}\right]^{\top}$ are interpreted as proxies for latent factors that drive the daily evolution of the IVS. \cite{goncalves2006predictable} subsequently proceed to show that modeling the dynamics of $\widehat{\alpha}_{t}$ can provide superior out-of-sample forecasts of IV.

One drawback of \possessivecite{goncalves2006predictable} specification is the assumed symmetry in the moneyness dimension of the IVS. A second drawback is that the model linearly approximates the term structure of the IVS. In the context of our study, commodity markets have been shown to exhibit both IV skews and non-linear term structure. To address these two issues, \cite{chalamandaris2011important} employ an augmented version of \possessivecite{dumas1998implied} parametric specification, namely:
\begin{align*}
\sigma_{i,t}=\ &\beta_{0,t}+\beta_{1,t}\mathbf{1}_{\left\{ \Delta_{i}>0\right\} }\Delta_{i}^{2}+\beta_{2,t}\mathbf{1}_{\left\{ \Delta_{i}<0\right\} }\Delta_{i}^{2}+\beta_{3,t}\frac{1-e^{-\lambda\tau_{i}}}{\lambda\tau_{i}}+\beta_{4,t}\left(\frac{1-e^{-\lambda\tau_{i}}}{\lambda\tau_{i}}-e^{-\lambda\tau_{i}}\right)+
\\
&\beta_{5,t}\mathbf{1}_{\left\{ \Delta_{i}>0\right\} }\Delta_{i}\tau_{i}+\beta_{6,t}\mathbf{1}_{\left\{ \Delta_{i}<0\right\} }\Delta_{i}\tau_{i}+\epsilon_{i,t},
\end{align*}
where $\mathbf{1}_{\left\{ x\right\} }$ is an indicator function that takes the value one if condition $x$ is true, and zero otherwise. The terms $\tfrac{(1-e^{-\lambda\tau_{i}})}{\lambda\tau_{i}}$ and $\frac{1-e^{-\lambda\tau_{i}}}{\lambda\tau_{i}}-e^{-\lambda\tau_{i}}$ represent the term structure of the IVS and are based on the Nelson-Siegel factors as successfully employed by \cite{diebold2006forecasting} to describe and forecast the yield curve. The $\lambda$ parameter determines the exponential decay rate of the term structure. In line with \cite{diebold2006forecasting}, we first utilize a non-linear estimation of all parameters, including $\lambda$, by minimizing the daily sum of squared errors. Following this we fix $\lambda$ to be equal to the median estimated value from the first step, and re-estimate the model using ordinary least squares. Coefficients $\beta_{1,t}$ and $\beta_{2,t}$ capture the IV skew by modeling in-the-money and out-of-the-money options separately. Similarly, the coefficients $\beta_{5,t}$ and $\beta_{6,t}$ capture the attenuation of the smile with time-to-maturity. \cite{chalamandaris2011important} show that modeling the dynamics of $\boldsymbol{\widehat{\beta}_{t}}=\left[\widehat{\beta}_{0,t},\widehat{\beta}_{1,t},\ldots,\widehat{\beta}_{6,t}\right]^{\top}$ leads to accurate forecasts of the entire IVS out-of-sample. The methodology for forecasting these coefficients is outlined Section \ref{subsec:Forecast Evaluation}.

We also consider a regression tree approach. A regression tree is a set of logical conditions that recursively creates a binary partition of the predictor space. The algorithm has three main components: (1) a way to select a split rule; (2) a rule to determine when a tree node is terminal; and (3) a standard for assigning a value to each terminal node. Its ability to split variables to construct a regression tree with end nodes is of great importance. 

In \cite{AC10}, the IV is regressed on a vector of predictors $\bm{x}^{\text{pred}}$ through an unspecified function $f_{m,\tau}$ such that
\begin{align*}
\sigma_{\Delta, \tau}^{\text{IV}} &= f_{\Delta,\tau}\left(\bm{x}^{\text{pred}}\right)+\epsilon_{\Delta,\tau} \\
&=f_{\Delta, \tau}\left(\Delta, \tau\right)+\epsilon_{\Delta, \tau},
\end{align*}
where $\Delta$ denotes option moneyness and $\tau$ denotes maturity. $\text{E}[\epsilon_{\Delta,\tau}] = 0$ and $\text{E}[\epsilon_{\Delta,\tau}^2]<\infty$ for each $\Delta, \tau>0$. The functions $f_{\Delta, \tau}(\cdot)$ are estimated by minimizing the expectation of a given loss function in a regression tree framework. However, when a decision tree is formed in this manner, the classification algorithm can generate some unwanted rules as it grows deeper, referred to as overfitting. To avoid this, we prune our regression tree using cost-complexity pruning as proposed by \cite{BFO+84} \citep[see][for further technical details]{EMS97}.

Following the approach of \cite{AC10}, we specify our regression function $f_{\Delta, \tau}$ to be a linear additive expansion of regression trees. More specifically, we use Gini's diversity index to choose the splits up to a maximum of ten end nodes. The complexity parameter is selected using the minimum cross-validated error calculated across the full in-sample period 2006-2014. This selected complexity parameter is then used as a constant throughout our forecasting exercise. Computationally, the regression trees are implemented using the \textit{rpart} package \citep{TA18} in R \citep{Team18}. Once the regression function $\widehat{f}_{\Delta, \tau}$ is estimated, the forecast $\sigma_{\Delta, \tau}^{\text{IV}}$ is obtained by providing it with the new set of moneyness and maturity variables. This forecasting model is henceforth referred to as RT.

\subsection{Time Series Properties of Parameters}\label{subsec:Forecast Evaluation}

First, we estimate the models of \cite{goncalves2006predictable} (GG) and \cite{chalamandaris2011important} (CT) on a daily basis. The average estimated coefficients over the in-sample period presented in Table~\ref{tab:Estimated_Coefficients}. Focusing first on the CT model, the skews in the moneyness dimension observed here are exhibited through the different coefficients for the left and right smiles. CT also estimates coefficients for both short- and medium-term structure variables, with negative estimates calculated for Gold options for instance. This relates to a dynamic that options with longer-dated expiries have higher IV than those with short-term maturity, and corresponds to the dynamics observed in Figure~\ref{fig:Average-Implied-Volatility}. It should be noted here, however, that for Cocoa options the majority of days exhibits insignificant parameter estimates.

\begin{table}[!htbp]
\tabcolsep 0.035in
\caption{Average Estimated Coefficients}\label{tab:Estimated_Coefficients}
\begin{centering}
\begin{tabular}{@{}lrrrrrrrrrrrrrrrrrr@{}}
\toprule
 & \multicolumn{5}{l}{\footnotesize \textit{\cite{goncalves2006predictable} estimation}} &  & \multicolumn{7}{l}{\footnotesize \textit{\cite{chalamandaris2011important} estimation}}\tabularnewline
\cline{2-6} \cline{8-14} 
\footnotesize{Commodity} & \textit{$\alpha_{0}$} & \textit{$\alpha_{1}$} & \textit{$\alpha_{2}$} & \textit{$\alpha_{3}$} & \textit{$\alpha_{4}$} &  & {$\beta_{0}$} & {$\beta_{1}$} & {$\beta_{2}$} & {$\beta_{3}$} & {$\beta_{4}$} & {$\beta_{5}$} & {$\beta_{6}$}\tabularnewline
\midrule 
\footnotesize{Cocoa} & 2.247 & -44.508 & 0.024 & 269.184 & -0.498 &  & 0.301 & 0.515 & 1560.918 & -0.011 & 0.154 & -0.074 & 5.326\tabularnewline
\footnotesize{Corn} & 0.330 & 0.022 & -0.003 & 0.332 & -0.001 &  & 0.254 & 0.856 & 0.728 & 0.063 & 0.123 & -0.008 & 0.014\tabularnewline
\footnotesize{Cotton} & 0.308 & 0.037 & -0.006 & 0.404 & -0.006 &  & 0.183 & 0.855 & 12.216 & 2.025 & -2.217 & -0.010 & 0.055\tabularnewline
\footnotesize{Soybean} & 0.268 & 0.054 & -0.001 & 0.601 & -0.007 &  & 0.237 & 1.655 & 0.819 & 53.163 & -53.236 & -0.011 & 0.008\tabularnewline
\footnotesize{Soybean Oil} & 0.204 & 0.315 & 0.015 & 1.313 & -0.179 &  & 0.242 & 2.246 & -193.687 & -7.619 & 10.879 & -0.044 & 0.069\tabularnewline
\footnotesize{Sugar} & 0.345 & 0.075 & -0.006 & 0.851 & -0.025 &  & 0.298 & 1.576 & -7.398 & 0.009 & 0.149 & -0.017 & -0.061\tabularnewline
\footnotesize{Wheat} & 0.344 & 0.049 & -0.005 & 0.521 & -0.005 &  & 0.286 & 1.006 & 0.898 & 75.931 & -78.678 & -0.006 & 0.003\tabularnewline
\footnotesize{Crude Oil} & 0.337 & -0.014 & -0.004 & 0.910 & -0.004 &  & 0.218 & 0.845 & 1.506 & 0.118 & 0.102 & -0.007 & 0.003\tabularnewline
\footnotesize{Heating Oil} & 0.269 & 0.097 & 0.002 & 0.599 & -0.024 &  & 0.283 & 0.562 & -210.339 & 0.001 & -0.047 & 0.017 & 0.244\tabularnewline
\footnotesize{Natural Gas} & 0.464 & -0.031 & -0.023 & 2.348 & 0.001 &  & 0.290 & -7.135 & 3.013 & 0.237 & 0.167 & -0.014 & 0.122\tabularnewline
\footnotesize{Gold} & 0.199 & 0.047 & 0.003 & 1.523 & -0.012 &  & 0.252 & 2.162 & 3.595 & -0.057 & -0.045 & -0.017 & 0.041\tabularnewline
\footnotesize{Silver} & 0.345 & 0.030 & 0.000 & 1.558 & -0.006 &  & 0.359 & 2.683 & -66.433 & -0.007 & -0.044 & -0.012 & -0.336\tabularnewline
\bottomrule
\end{tabular}
\par\end{centering}
\centering{}{\footnotesize For each of the 12 different commodity options in our sample, this table presents the average daily estimated coefficients for the \cite{goncalves2006predictable} and \cite{chalamandaris2011important} models fitted to our implied volatility data set during the January 2006-December 2014 in-sample period.}{\footnotesize \par}
\end{table}

\begin{figure}[!htbp]
\caption{Gold Commodity Options; Estimated GG-Coefficients (2010)}\label{fig:GCA-GG-Coefficients}
\begin{centering}
\includegraphics[width = 15.6cm]{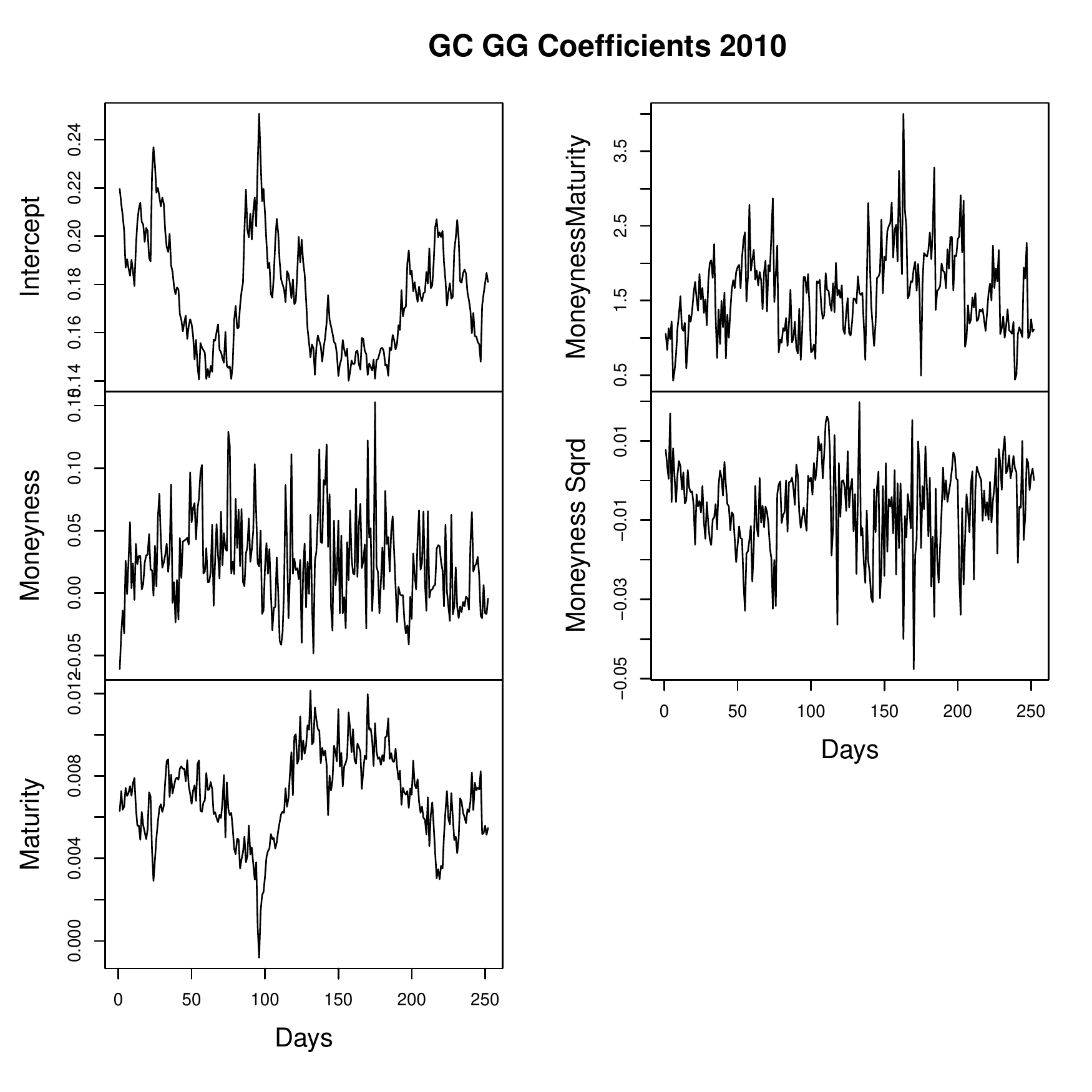}
\par\end{centering}
\noindent \centering{}{\footnotesize{}This figure presents the time variation of the five coefficient values of the \cite{goncalves2006predictable} model when estimated for Gold commodity options during the year 2010. The coefficients presented refer to the variables of intercept, moneyness, maturity, moneyness multiplied by maturity, and moneyness squared.}{\footnotesize \par}
\end{figure}

\begin{figure}[!htbp]
\begin{centering}
\caption{Gold Commodity Options; Estimated CT-Coefficients (2010)}\label{fig:GCA-CT-Coefficients}
\includegraphics[width=15.6cm]{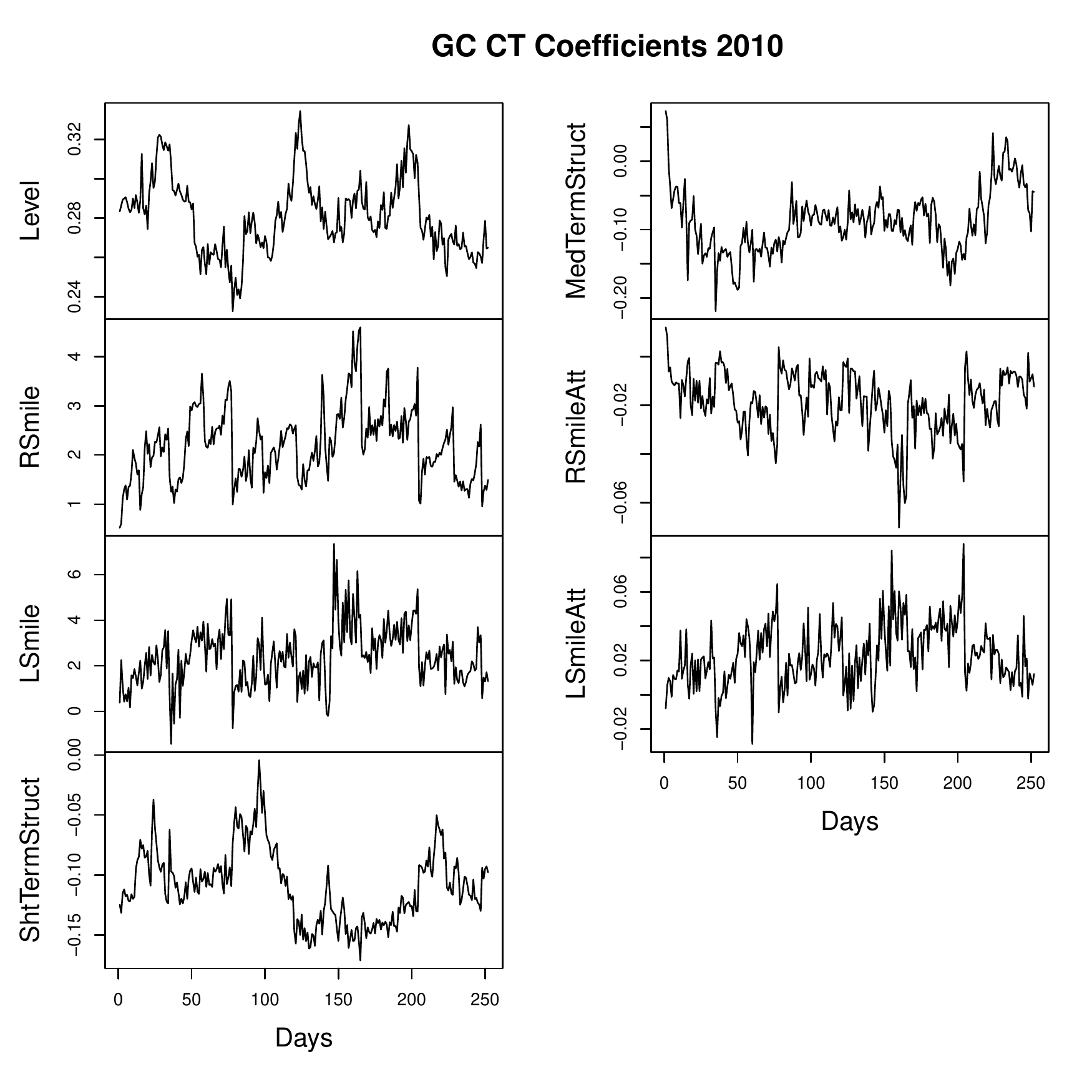}
\par\end{centering}
\noindent \centering{}{\footnotesize{}This figure presents the time variation of the seven coefficient values of the \cite{chalamandaris2011important} model when estimated for Gold commodity options during the year 2010. The coefficients presented refer to the variables of implied volatility level, right smile, left smile, short-term structure, medium-term structure, right smile attenuation and left smile attenuation.}{\footnotesize \par}
\end{figure}

\begin{figure}[!htbp]
\begin{centering}
\caption{Autocorrelation Function of GG-Coefficients for Gold Commodity Options}\label{fig:GG-ACF}
\includegraphics[width = \textwidth]{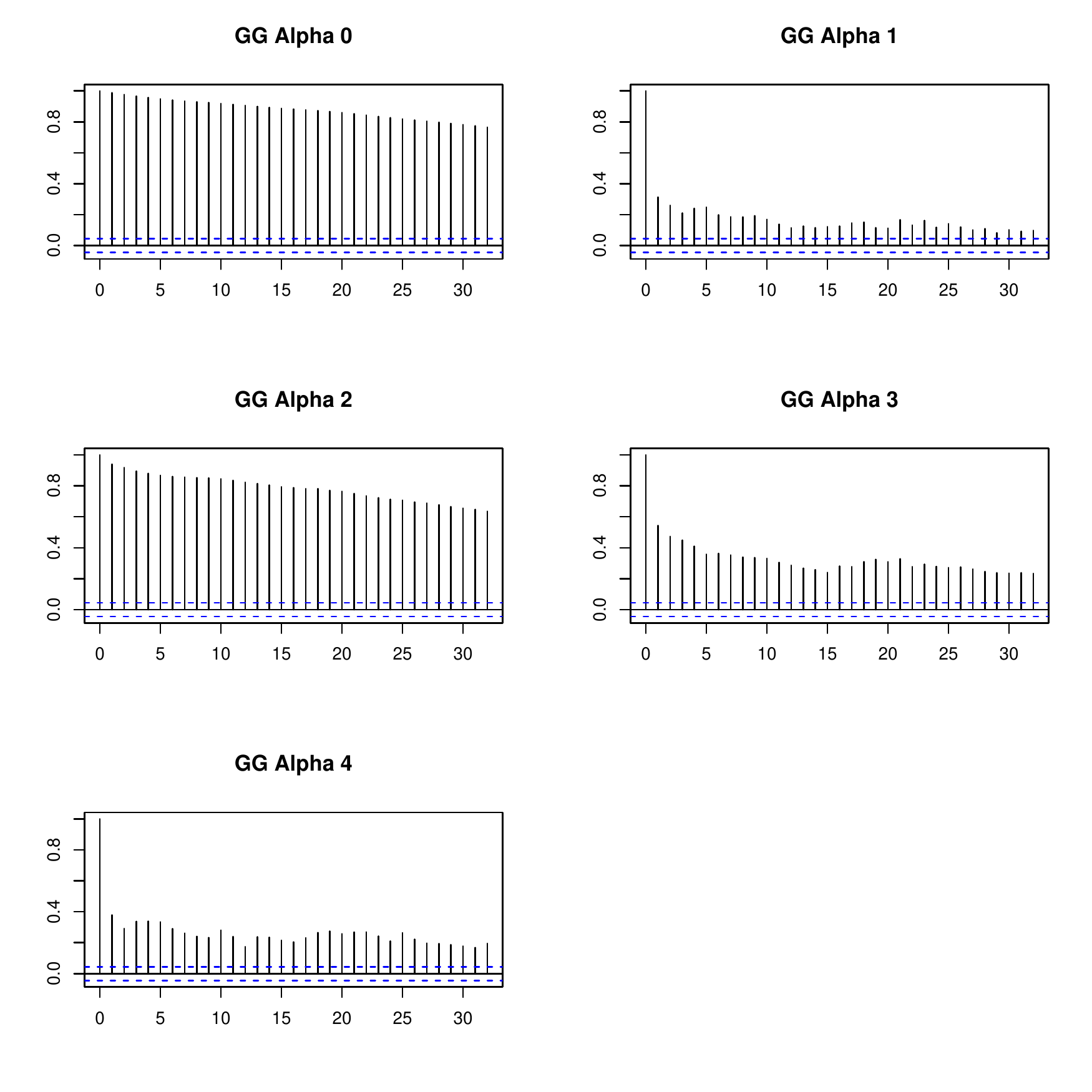}
\par\end{centering}
\noindent \centering{}{\footnotesize{}This figure presents the calculated autocorrelation function for the five coefficient values of the \cite{goncalves2006predictable} model when estimated for Gold commodity options during the in-sample January 2006-December 2014 period. The $x$-axis represents the lags in days. The Alpha coefficients presented refer to the variables of intercept, moneyness, maturity, moneyness multiplied by maturity, and moneyness squared.}{\footnotesize \par}
\end{figure}

\begin{figure}[!htbp]
\begin{centering}
\caption{Autocorrelation Function of CT-Coefficients for Gold Commodity Options}\label{fig:CT-ACF}
\includegraphics[width=15.6cm]{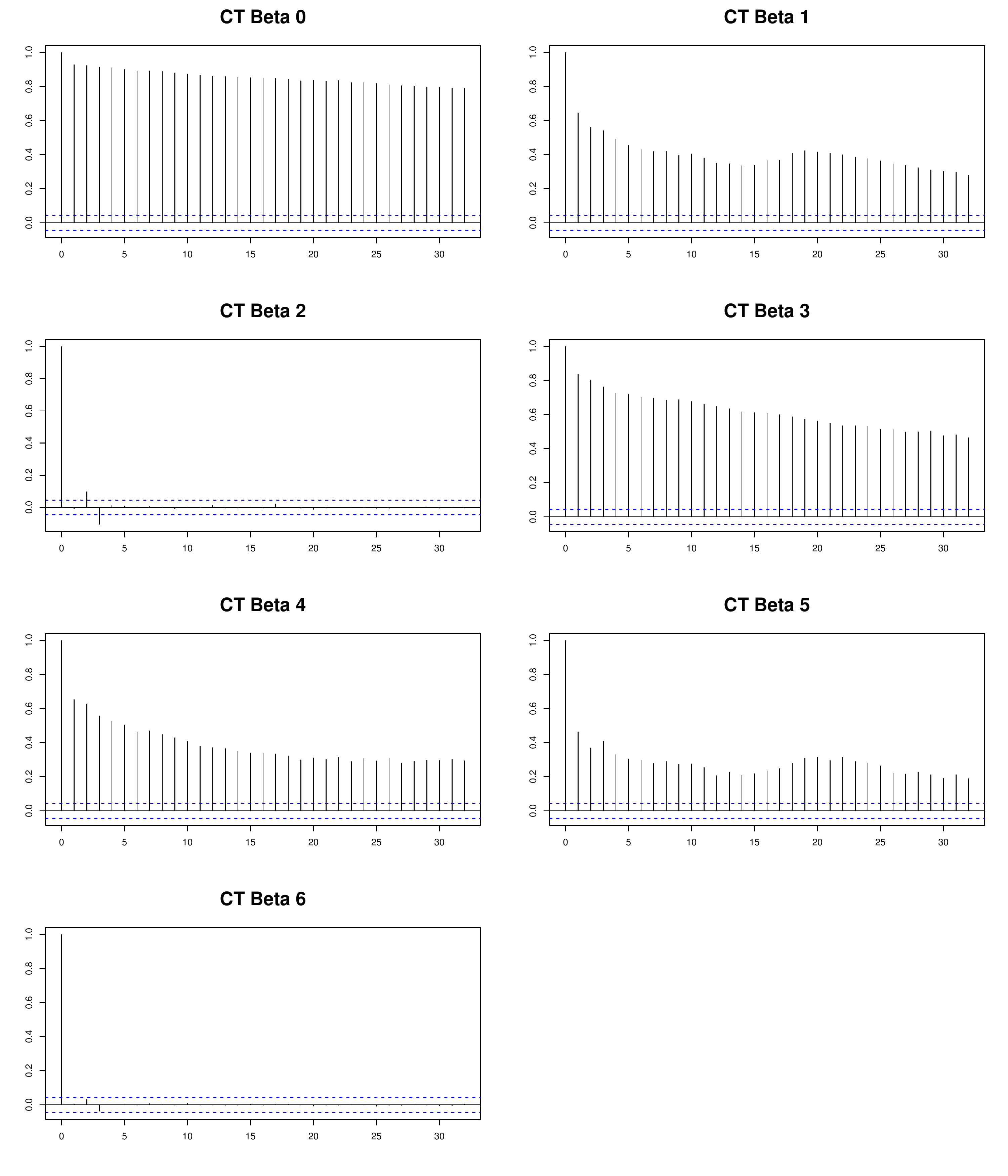}
\par\end{centering}
\noindent \centering{}{\footnotesize{}This figure presents the calculated autocorrelation function for the seven coefficient values of the \cite{chalamandaris2011important} model when estimated for Gold commodity options during the in-sample January 2006-December 2014 period. The x-axis represents the lags in days. The Beta coefficients presented refer to the variables of implied volatility level, right smile, left smile, short-term structure, medium-term structure, right smile attenuation, and left smile attenuation.}{\footnotesize \par}
\end{figure}

\begin{table}[!htbp]
\centering
\tabcolsep 0.2in
\caption{Correlation Matrix of GG-Coefficients}\label{GG-corr-matrix}
\begin{tabular}{@{}llllll@{}}
\toprule
                  & $\alpha_{0}$ & $\alpha_{1}$ & $\alpha_{2}$ & $\alpha_{3}$ & $\alpha_{4}$ \\
                  \midrule
$\alpha_{0}$         & 1		  & -0.262***    & -0.536***   & -0.362***            & 0.375***          \\
$\alpha_{1}$         & -0.262***    & 1		  & 0.208***    & -0.211***            & -0.743***        \\
$\alpha_{2}$          & -0.536***    & 0.208***     & 1		 & 0.082***             & -0.265***         \\
$\alpha_{3}$ & -0.362***    & -0.211***    & 0.082***    & 1	             & -0.301***         \\
$\alpha_{4}$    & 0.375***     & -0.743***    & -0.265***   & -0.301***            & 1         \tabularnewline
\bottomrule
\end{tabular}
\noindent \centering{}{\footnotesize{}\\This table presents the calculated correlation matrix between the five coefficient values of the \cite{goncalves2006predictable} model when estimated for the Gold commodity option during the in-sample January 2006-December 2014 period. The alpha coefficients, $\alpha_{0}$, $\alpha_{1}$, $\alpha_{2}$, $\alpha_{3}$ and $\alpha_{4}$, refer to the variables of intercept, moneyness, maturity, moneyness multiplied by maturity, and moneyness squared, respectively. *, **, and ***, indicate significance at 10\%, 5\%, and 1\%, respectively.}{\footnotesize \par}
\end{table}

\begin{table}[!htbp]
\centering
\tabcolsep 0.2in
\caption{Correlation Matrix of CT-Coefficients}\label{CT-corr-matrix}
\begin{tabular}{@{}llllllll@{}}
\toprule
              & $\beta_{0}$  & $\beta_{1}$ & $\beta_{2}$ & $\beta_{3}$ & $\beta_{4}$ & $\beta_{5}$ & $\beta_{6}$ \\
              \midrule
$\beta_{0}$         & 1      & -0.227*** & 0.064***  & -0.169***        & -0.115***        & 0.058**     & 0.056**     \\
$\beta_{1}$        & -0.227*** & 1      & 0.036  & -0.125***        & -0.350***         & -0.864***    & 0.029     \\
$\beta_{2}$        & 0.064***  & 0.036  & 1      & -0.032        & -0.045**        & -0.057**    & 0.984***     \\
$\beta_{3}$ & -0.169*** & -0.125*** & -0.032 & 1             & -0.13***         & 0.123***     & -0.026    \\
$\beta_{4}$ & -0.115*** & -0.350***  & -0.045** & -0.13***         & 1             & 0.453***     & -0.037    \\
$\beta_{5}$     & 0.058**  & -0.864*** & -0.057** & 0.123***         & 0.453***         & 1         & -0.043*    \\
$\beta_{6}$     & 0.056**  & 0.029  & 0.984***  & -0.026        & -0.037        & -0.043*    & 1     \tabularnewline
\bottomrule   
\end{tabular}
\noindent \centering{}{\footnotesize{}This table presents the calculated correlation matrix between the seven coefficient values of the \cite{chalamandaris2011important} model when estimated for the Gold commodity option during the in-sample January 2006-December 2014 period. The beta coefficients, $\beta_{0}$, $\beta_{1}$, $\beta_{2}$, $\beta_{3}$, $\beta_{4}$, $\beta_{5}$ and $\beta_{6}$, refer to the variables of implied volatility level, right smile, left smile, short-term structure, medium-term structure, right smile attenuation, and left smile attenuation, respectively. *, **, and ***, indicate significance at 10\%, 5\%, and 1\%, respectively.}{\footnotesize \par}
\end{table}

Estimating GG and CT models on a daily basis, as we do, produces sets of dynamic factors, $\widehat{\boldsymbol{\alpha}}_{t}$ and $\bm{\widehat{\beta}}_{\boldsymbol{t}}$, respectively. Figures \ref{fig:GCA-GG-Coefficients} and \ref{fig:GCA-CT-Coefficients} show the evolution of the coefficients estimated throughout 2010. They demonstrate that the coefficients are not constant over time, with some coefficients even alternating between positive and negative states throughout a single year. We now empirically investigate the time series properties of these parameters. This, in turn, informs our choice of which time series processes to adopt to model the evolution of the factors. In order to establish the time series properties we present the cross-correlation matrices for GG and CT separately in Tables \ref{GG-corr-matrix} and \ref{CT-corr-matrix}, and the autocorrelation functions of each coefficient in Figures \ref{fig:GG-ACF} and \ref{fig:CT-ACF}.

First, we focus on the autocorrelation function plots to identify temporal dependencies within each parameter series. There is a pattern of high autocorrelation function (ACF) values observed across the majority of the series, motivating the adoption of a simple autoregressive (AR) process to model this evolution. Second, parameters such as Beta 2 and Beta 6 of the CT model may be well modeled by a moving average model, whereby the ACF values decay dramatically at low lag orders. We incorporate this dynamic along with the widespread autoregressive dependency established previously through the Autoregressive Integrated Moving Average (ARIMA). In the ARIMA, $d$ is selected based on successive Kwiatkowski-Phillips-Schmidt-Shin (KPSS) unit root tests \citep{KPSS92}.\footnote{KPSS tests are primarily used for testing the null hypothesis that an observable time series is stationary around a deterministic trend, although KPSS tests can also detect the presence of a long-memory process.} We test the original data for a unit root; if the test result is significant, then we test the differenced data for a unit root. The procedure continues until we obtain our first insignificant result. Third, based on the possible exponential decay patterns observed across GG Alpha 1, CT Beta 3, and CT Beta 4, we consider an exponential smoothing time series approach with optimal trend and seasonal components selected using a corrected Akaike information criterion. Finally, we turn our attention to Tables \ref{GG-corr-matrix} and \ref{CT-corr-matrix}, where we learn from the cross-correlation matrices that significant linear relationships exist between the coefficient estimates produced by both models. This motivates the use of a tractable multivariate time series approach to capture these interacting dynamics simultaneously; for this reason, we also specify a VAR model. We now provide more details of these time series models.

In line with \cite{Chalamandaris2014} we hypothesize that the shape of the IVS at time $t+h$ depends on factors $\widehat{\boldsymbol{\alpha}}_{t+h}$ and $\widehat{\boldsymbol{\beta}}_{t+h}$. To model the time evolution of $\widehat{\boldsymbol{\alpha}}_{t},$ and $\widehat{\boldsymbol{\beta}}_{t}$ we employ various time series models, namely AR, VAR, ARIMA, and Exponential Smoothing (ETS), to produce forecasts of their coefficients. Finally, we also adopt a no change driftless random walk (RW) model to act as an intuitive benchmark.

More specifically, forecasts are produced using:
\[
\widehat{\sigma}_{i,t+h}=\widehat{\alpha}_{0,t+h}+\widehat{\alpha}_{1,t+h}\Delta_{i}+\widehat{\alpha}_{2,t+h}\Delta_{i}^{2}+\widehat{\alpha}_{3,t+h}\tau_{i}+\widehat{\alpha}_{4,t+h}\left(\Delta_{i}\times\tau_{i}\right)+\widehat{\epsilon}_{i,t+h},
\]
and
\begin{align*}
\widehat{\sigma}_{i,t+h}=\ & \widehat{\beta}_{0,t+h}+\widehat{\beta}_{1,t+h}\mathbf{1}_{\left\{ \Delta_{i}>0\right\} }\Delta_{i}^{2}+\widehat{\beta}_{2,t+h}\mathbf{1}_{\left\{ \Delta_{i}<0\right\} }\Delta_{i}^{2}+\widehat{\beta}_{3,t+h}\frac{1-e^{-\lambda\tau_{i}}}{\lambda\tau_{i}}+\widehat{\beta}_{4,t+h}\left(\frac{1-e^{-\lambda\tau_{i}}}{\lambda\tau_{i}}-e^{-\lambda\tau_{i}}\right)+\\
&\widehat{\beta}_{5,t+h}\mathbf{1}_{\left\{ \Delta_{i}>0\right\} }\Delta_{i}\tau_{i}+\widehat{\beta}_{6,t+h}\mathbf{1}_{\left\{ \Delta_{i}<0\right\} }\Delta_{i}^{2}+\widehat{\epsilon}_{i,t+h},
\end{align*}
where $\widehat{\boldsymbol{\alpha}}_{t+h}=\left[\widehat{\alpha}_{0,t+h},\ldots,\widehat{\alpha}_{4,t+h}\right]^{\top}$ and $\widehat{\boldsymbol{\beta}}_{t+h}=\left[\widehat{\beta}_{0,t+h},\ldots,\widehat{\beta}_{6,t+h}\right]^{\top}$ are the result of time-series specifications:
\begin{itemize}
\item AR (henceforth referred to as GG-AR and CT-AR, for the \cite{goncalves2006predictable} and \cite{chalamandaris2011important} models, respectively)
\item ARIMA (henceforth referred to as GG-ARIMA and CT-ARIMA, for the \cite{goncalves2006predictable} and \cite{chalamandaris2011important} models, respectively)
\item ETS (henceforth referred to as GG-ETS and CT-ETS, for the \cite{goncalves2006predictable} and \cite{chalamandaris2011important} models, respectively)
\item VAR (henceforth referred to as GG-VAR and CT-VAR, for the \cite{goncalves2006predictable} and \cite{chalamandaris2011important} models, respectively).
\item As a no change benchmark we also specify the RW (henceforth referred to as GG-RW and CT-RW, for the \cite{goncalves2006predictable} and \cite{chalamandaris2011important} models, respectively). This essentially amounts to keeping the parameters constant.
\end{itemize}
Forecasting horizons of 1, 2, 5, 10, and 30 days ahead are considered. For forecasts of longer than one-day ahead we implement direct multi-day ahead forecasts. Our choice of direct forecasts is based on \cite{Cox61}, who concludes that they are more efficient than iterated forecasts in the context of exponential smoothing, and \cite{Klein68}, who suggest employing direct multiperiod estimation of dynamic forecasting models. Further, \cite{Bhansali99} and \cite{Ing03} conclude that the robustness of the direct approach to model misspecification makes it a more attractive procedure than the bias-prone indirect approach.

\subsection{Out-of-Sample Forecasting}\label{sec:Out of Sample Forecasting}

We assess the forecast performance of each model using the following measures:
\begin{enumerate}
\item Root mean squared error (RMSE) is a measure of the differences between the realized values and the values predicted by a model. It is defined as the square root of the mean squared forecast error and serves to aggregate the errors into a single measure of predictive power:\\
\[
\text{RMSE}=\sqrt{\frac{\sum_{i=1}^{n}(\sigma_{i,t+h}-\widehat{\sigma}_{i,t+h})^{2}}{n}},
\]
where $\sigma_{i,t+h}$ are the observed values and $\widehat{\sigma}_{i,t+h}$ are the values predicted from the model.
\item Root mean squared percentage error (RMSPE) is also a measure of the differences between the realized values and the values predicted by a model, however it is scaled by the realized observations and is expressed in percentage terms, as follows:\\
\[
\text{RMSPE}=\sqrt{\frac{\sum_{i=1}^{n}\Big[\frac{100(\sigma_{i,t+h}-\widehat{\sigma}_{i,t+h})}{\sigma_{i,t+h}}\Big]^{2}}{n}},
\]
where $\sigma_{i,t+h}$ are the observed values and $\widehat{\sigma}_{i,t+h}$ are the values predicted from the model.
\item Mean absolute percentage error (MAPE) is the average of the absolute differences between the realized values and the values predicted by the model, again scaled by the realized observations and expressed in percentage terms:\\
\[
\text{MAPE}=\frac{100}{n}{\displaystyle \sum_{i=1}^{n}\left|\frac{\sigma_{i,t+h}-\widehat{\sigma}_{i,t+h}}{\sigma_{i,t+h}}\right|,}
\]
where $\sigma_{i,t+h}$ are the observed values and $\widehat{\sigma}_{i,t+h}$ are the values predicted from the model.
\item The sign success ratio (SSR) is the percentage of predictions for which the change in the values predicted by the model, $\widehat{\sigma}_{i,t+h}$, have the same sign as the corresponding change in the realized values, $\sigma_{i,t+h}$. The SSR measures how well the model can predict the direction of movement, regardless of error magnitude.\\
\end{enumerate}

We implement a rolling window approach to produce forecasts for each of our 500 out-of-sample days. Using a size-constant rolling window of 1,167 days, we produce daily forecasts. The forecasts we produce are based only on values available at the date on which the forecast is made. Further, we never expand the window by adding new observations. Instead, we add the new observation and remove the oldest one when rolling over. This rolling window approach has the advantage of mitigating the impact of structural breaks. As a further sensibility check, we constrain all time-series IV forecasts to reside in the $[0\%,100\%]$ range. We formally test if any outperformance uncovered is sample specific or if we can draw inferences regarding the entire population. To this end, we utilize the Model Confidence Set outlined in the next section to establish which model(s) are statistically superior.

\subsection{Model Confidence Set}

To examine statistical significance among our multiple approaches, we employ the model confidence set (MCS) procedure. The MCS procedure proposed by \cite{HLN11} consists of a sequence of tests that permit construction of a set of ``superior" models, where the null hypothesis of equal predictive ability (EPA) is not rejected at a specified confidence level. As the EPA test statistic can be evaluated for any loss function, we follow \cite{HLN11} and adopt the popular squared error measure at a 75\% confidence level.

The procedure begins with our initial set of models of dimension $m=11$ encompassing all the IVS forecasting models considered, $M_0 = \{M_1, M_2, M_3, M_4, M_5, M_6, M_7, M_8, M_9, M_{10}, M_{11}\}$. For a given confidence level, the superior set of models $\widehat{M}_{1-\alpha}^*$, is determined where $m^*\leq m$. The best scenario is when the final set consists of a single model that is, $m=1$. First, let $d_{ij,t}$ denote the loss differential between models $i$ and $j$, that is:
\begin{equation*}
d_{ij,t} = l_{i,t} - l_{j,t}, \qquad i,j=1,\dots,m, \quad t=1,\dots,n.
\end{equation*} 
The EPA hypothesis for a given set of $\text{M}$ candidate models can subsequently be formulated by:
\begin{align}
\text{H}_{\text{0,M}}: c_{ij}&=0, \qquad \text{for all}\quad i, j = 1,2,\dots,m\notag\\
\text{H}_{\text{A,M}}: c_{ij}&\neq 0, \qquad \text{for some}\quad i, j = 1,2,\dots,m,\label{eq:hypo_1}
\end{align}
where $c_{ij} = \text{E}(d_{ij})$ is assumed to be finite and time independent. Based on $c_{ij}$ we construct a hypothesis test as follows:
\begin{equation}
t_{ij} = \frac{\overline{d}_{ij}}{\sqrt{\widehat{\text{Var}}\left(\overline{d}_{ij}\right)}}, \label{eq:t2} 
\end{equation}
where $\overline{d}_{ij} =\frac{1}{n}\sum^n_{t=1}d_{ij,t}$ measures the relative sample loss between the $i^{\text{th}}$ and $j^{\text{th}}$ models. Note that $\widehat{\text{Var}}\left(\overline{d}_{ij}\right)$ are the bootstrapped estimates of $\text{Var}\left(\overline{d}_{ij}\right)$. To produce these we perform a block bootstrap procedure with 5,000 bootstrap samples, based on \cite{HLN11} and \cite{BC15}, where the block length $p$ is given by the maximum number of significant parameters obtained by fitting an AR($p$) process to all the $d_{ij}$ terms. For the hypothesis in~\eqref{eq:hypo_1}, we utilize the test statistic:
\begin{equation}
T_{\text{R,M}} = \max_{i,j\in \text{M}}|t_{ij}|,\label{eq:MCS_1}
\end{equation} 
where $t_{ij}$ is defined in~\eqref{eq:t2}.
 
MCS is a sequential testing procedure that eliminates the worst model at each step until the hypothesis of EPA is accepted for all the models belonging to a set of superior models. The selection of the worst model is determined by an elimination rule that is consistent with the test statistic:
\begin{equation*}
e_{\text{R,M}} = \argmax_{i\in M}\left\{\sup_{j\in M}\frac{\overline{d}_{ij}}{\sqrt{\widehat{\text{Var}}\left(\overline{d}_{ij}\right)}}\right\}.
\end{equation*}

To summarize, the MCS procedure to obtain a superior set of models consists of the following steps:
\begin{enumerate}
\item[1)] Set $M=M_0$.
\item[2)] If the null hypothesis is accepted, then $M^* = M$; otherwise use the elimination rules defined in~\eqref{eq:MCS_1} to determine the worst model.
\item[3)] Remove the worst model, and go to Step 2).
\end{enumerate}

\section{Out-of-Sample Results}\label{sec:Empirical Findings}
We now move to an out-of-sample forecasting environment. We employ the RMSPE and MAPE measures outlined in Section~\ref{sec:Out of Sample Forecasting} of the full IVS for all 12 commodities in our sample. Results are presented in Tables~\ref{tab:OOS_RMSPE} and \ref{tab:OOS_MAPE}, respectively. The first notable item is the large error size of the GG framework in this out-of-sample setting, a dynamic that is broadly consistent across the various commodity options and forecasting horizons. When we focus on the relative performances of the RT and CT models we note that the RT model leads to lower errors than the CT models for a number of Agricultural commodities; Cocoa, Cotton, and Soybean Oil. This observed dynamic holds at both short- and long-term forecasting horizons. In contrast, the Precious Metals in our sample exhibit lower forecasting errors using CT models; for example, the RMSPE for one-day-ahead Gold options forecasts of 14.14 and 5.69 for RT and CT-VAR, respectively.

In modeling the parameter estimates produced by the CT framework, a dynamic we observe is that, for a number of commodities, the VAR specification leads to lower RMSPEs and MAPEs at short forecasting horizons, with the simpler univariate AR specification producing lower errors at longer forecasting horizons. Examples of this are Natural Gas, Gold, and Silver commodity options. Focusing on Natural Gas for instance, we observe CT-AR and CT-VAR RMSPEs of 11.45 and 6.54, respectively, at a one-day-ahead forecasting horizon, in contrast with 11.37 and 12.90, respectively, at the 30-day-ahead horizon.\footnote{Sensitivity to other forecasting horizons is explored with broadly similar results uncovered for h=2- and 10-day-ahead forecasting horizons. For brevity these results are excluded.} 

\begin{table}[!htbp]
\tabcolsep 0.03in
{\caption{Out-of-Sample RMSPE}\label{tab:OOS_RMSPE}
}{\footnotesize \par}
\begin{centering}
\begin{tabular}{@{}lcccccccccccccc@{}}
\toprule
{\footnotesize Forecasting} & \multicolumn{7}{c}{{Agricultural}} &  & \multicolumn{3}{c}{{Energy}} &  & \multicolumn{2}{c}{{Metals}}\tabularnewline
\cline{2-8} \cline{10-12} \cline{14-15} 
{\footnotesize Horizon (h)} & {\footnotesize Cocoa} & {\footnotesize Corn} & {\footnotesize Cotton} & {\footnotesize Soybean} & {\footnotesize Soybean} & {\footnotesize Sugar} & {\footnotesize Wheat} &  & {\footnotesize Crude} & {\footnotesize Heating} & {\footnotesize Natural} &  & {\footnotesize Gold} & {\footnotesize Silver}\tabularnewline
& & & & & {\footnotesize Oil} & & & &{\footnotesize Oil} & {\footnotesize Oil}  & {\footnotesize Gas} \\
\midrule
{$h=1$}      &        &        &        &         &             &       &       &  &           &             &             &  &        &         \\
RT       & \textBF{20.30}& 25.55  & \textBF{29.65}& 23.77   & \textBF{10.81}       & 19.08 & 16.36 &  & 28.72     & 30.57       & 13.32       &  & 14.14  & 15.11   \\
GG-AR    & 335.31 & 103.10 & 54.06  & 128.90  & 110.08      & 55.05 & 69.96 &  & 55.39     & 39.33       & 59.34       &  & 167.43 & 78.49   \\
CT-AR    & 103.33 & 6.49   & 94.29  & 7.14    & 28.15       & 10.00 & \textBF{6.48}  &  & 8.19      & 40.86       & 11.45       &  & 5.88   & 6.35    \\
GG-VAR   & 284.40 & 105.84 & 60.50  & 135.69  & 113.86      & 48.21 & 72.69 &  & 47.59     & 49.13       & 57.13       &  & 170.22 & 87.21   \\
CT-VAR   & 110.98 & \textBF{6.44} & 112.33 & \textBF{6.96}  & 27.22       & 12.73 & 6.78  &  & \textBF{5.92}      & 48.38       & \textBF{6.54}        &  & \textBF{5.69}   & 4.97    \\
GG-ARIMA & 100.00 & 112.70 & 54.10  & 134.60  & 100.24      & 41.43 & 82.18 &  & 48.81     & \textBF{24.39}       & 58.68       &  & 164.10 & 79.81   \\
CT-ARIMA & 216.75 & 6.49   & 147.31 & 7.11    & 34.62       & \textBF{8.51}  & 6.71  &  & 6.71      & 46.51       & 8.75        &  & 5.77   & 5.07    \\
GG-ETS   & 342.83 & 112.84 & 55.47  & 136.52  & 110.04      & 41.39 & 83.05 &  & 49.32     & 42.63       & 61.37       &  & 162.78 & 80.38   \\
CT-ETS   & 152.35 & 6.45   & 116.03 & 7.07    & 26.31       & 8.71  & 6.69  &  & 6.73      & 36.09       & 8.75        &  & 5.70   & \textBF{4.90}    \\
         &        &        &        &         &             &       &       &  &           &             &             &  &        &         \\
{$h=5$}      &        &        &        &         &             &       &       &  &           &             &             &  &        &         \\
RT       & \textBF{20.30}  & 25.55  & \textBF{29.65}  & 23.77   & \textBF{10.81}   & 19.08 & 16.36 &  & 28.72     & \textBF{30.57}       & 13.32       &  & 14.14  & 15.11   \\
GG-AR    & 329.22 & 102.04 & 82.06  & 116.94  & 86.19       & 81.86 & 88.93 &  & 94.98     & 89.06       & 93.10       &  & 132.89 & 78.93   \\
CT-AR    & 103.38 & 9.34   & 94.43  & 10.92   & 28.35       & 10.12 & 8.97  &  & 9.74      & 40.14       & 9.01        &  & \textBF{8.06}   & 6.77    \\
GG-VAR   & 249.33 & 104.53 & 56.96  & 134.23  & 113.38      & 50.72 & 67.47 &  & 46.36     & 51.10       & 57.76       &  & 173.97 & 85.25   \\
CT-VAR   & 112.95 & 9.27   & 111.05 & \textBF{9.67}    & 31.88       & 10.57 & \textBF{8.84}&  & \textBF{9.04}  & 45.75       & \textBF{7.73}        &  & 8.35   & 7.01    \\
GG-ARIMA & 100.00 & 107.71 & 81.51  & 121.75  & 84.19       & 87.87 & 90.78 &  & 96.11     & 91.47       & 93.85       &  & 132.54 & 79.34   \\
CT-ARIMA & 214.30 & 9.63   & 147.38 & 10.88   & 34.73       & 9.71  & 9.27  &  & 9.64      & 45.63       & 9.64        &  & 8.30   & \textBF{6.42}   \\
GG-ETS   & 312.13 & 107.98 & 81.31  & 123.53  & 85.77       & 87.95 & 91.26 &  & 95.94     & 90.99       & 94.56       &  & 132.24 & 79.56   \\
CT-ETS   & 154.46 & \textBF{9.22}   & 114.52 & 9.69    & 27.04       & \textBF{9.28} & 8.96  &  & 9.56      & 35.41       & 9.61        &  & 8.16   & 6.56    \\
         &        &        &        &         &             &       &       &  &           &             &             &  &        &         \\
{$h=30$}     &        &        &        &         &             &       &       &  &           &             &             &  &        &         \\
RT       & \textBF{20.30}  & 25.55  & \textBF{29.65}  & 23.77   & \textBF{10.81} & 19.08 & 16.36 &  & 28.72     & \textBF{30.57}       & 13.32       &  & 14.14  & 15.11   \\
GG-AR    & 329.22 & 99.05  & 82.14  & 112.00  & 86.18       & 81.83 & 88.29 &  & 94.21     & 89.64       & 93.31       &  & 134.64 & 78.90   \\
CT-AR    & 103.70 & 20.34  & 93.54  & 21.84   & 28.62       & 12.82 & 15.48 &  & \textBF{16.21}     & 38.91       & \textBF{11.37}     &  & \textBF{11.23}  & \textBF{9.42}    \\
GG-VAR   & 239.82 & 89.18  & 70.64  & 126.52  & 106.13      & 59.96 & 62.41 &  & 53.19     & 66.30       & 60.75       &  & 177.72 & 85.95   \\
CT-VAR   & 146.90 & \textBF{17.46}  & 150.34 & \textBF{15.88}   & 62.49       & 22.70 & 18.98 &  & 21.75     & 72.19       & 12.90       &  & 14.11  & 11.30   \\
GG-ARIMA & 100.00 & 103.78 & 81.66  & 115.78  & 84.61       & 86.89 & 90.15 &  & 95.18     & 91.35       & 94.81       &  & 134.74 & 79.50   \\
CT-ARIMA & 207.98 & 21.85  & 146.53 & 22.64   & 36.24       & 15.31 & 16.73 &  & 18.60     & 40.17       & 15.02       &  & 12.53  & 9.62    \\
GG-ETS   & 312.09 & 104.06 & 81.43  & 117.56  & 85.38       & 87.14 & 90.71 &  & 94.94     & 91.90       & 95.39       &  & 134.58 & 79.68   \\
CT-ETS   & 168.61 & 18.47  & 132.03 & 19.82   & 28.37       & \textBF{12.48} & \textBF{12.85} &  & 18.46 & 33.13       & 13.46       &  & 12.86  & 9.95            \tabularnewline      
\bottomrule
\end{tabular}
\par\end{centering}{\tiny \par}
\centering{}{\footnotesize This table presents the Root Mean Squared Percentage Error (RMSPE) metric aggregated across all moneyness and term structure data points available for the 12 commodity options in our sample during the 500-day out-of-sample period (December 2014-December 2016). Results are presented for 1-, 5-, and 30-day-ahead forecasts. The lowest forecast errors for each commodity and forecast horizon are shown in bold.}{\footnotesize \par}
\end{table}
{\footnotesize \par}

\begin{table}[!htbp]
\tabcolsep 0.031in
{\caption{Out-of-Sample MAPE}\label{tab:OOS_MAPE}}{\footnotesize \par}
\begin{centering}
\begin{tabular}{@{}lcccccccccccccc@{}}
\toprule
{\footnotesize Forecasting} & \multicolumn{7}{c}{{Agricultural}} &  & \multicolumn{3}{c}{{Energy}} &  & \multicolumn{2}{c}{{Metals}}\tabularnewline
\cline{2-8} \cline{10-12} \cline{14-15} 
{\footnotesize Horizon (h)} & {\footnotesize Cocoa} & {\footnotesize Corn} & {\footnotesize Cotton} & {\footnotesize Soybean} & {\footnotesize Soybean} & {\footnotesize Sugar} & {\footnotesize Wheat} &  & {\footnotesize Crude} & {\footnotesize Heating} & {\footnotesize Natural} &  & {\footnotesize Gold}& {\footnotesize Silver}\tabularnewline
& & & & & {\footnotesize Oil} & & &  & {\footnotesize Oil} & {\footnotesize Oil} & {\footnotesize Gas} \\
\midrule
{$h=1$} &  &  &  &  &  &  &  &  &  &  &   &  &  & \tabularnewline
RT           & \textBF{15.18}  & 19.37 & \textBF{24.93}  & 19.08   & \textBF{8.57}    & 16.83 & 12.49 &  & 26.21     & 28.58       & 10.65       &  & 11.39  & 12.17   \\
GG-AR        & 319.33 & 68.12 & 36.25  & 82.07   & 78.28       & 39.10 & 43.32 &  & 30.26     & 22.49       & 31.35       &  & 120.15 & 56.76   \\
CT-AR        & 65.14  & 4.72  & 68.68  & 4.98    & 15.47       & 8.08  & 4.86  &  & 6.41      & 34.18       & 8.42        &  & 4.47   & 4.66    \\
GG-VAR       & 255.82 & 70.15 & 38.86  & 86.65   & 80.63       & 31.38 & 45.59 &  & 23.74     & 23.33       & 30.13       &  & 122.45 & 62.22   \\
CT-VAR       & 66.56  & \textBF{4.71}  & 82.79  & \textBF{4.66}  & 11.15     & \textBF{5.71} & \textBF{4.85}&  & \textBF{4.06}      & 27.73       & \textBF{4.56}        &  & \textBF{4.29}   & \textBF{3.58}    \\
GG-ARIMA     & 100.00 & 75.54 & 36.10  & 85.94   & 70.87       & 25.86 & 52.20 &  & 24.94     & \textBF{16.65}       & 30.46       &  & 115.38 & 57.69   \\
CT-ARIMA     & 173.39 & 4.77  & 143.36 & 4.79    & 17.78       & 6.13  & 5.05  &  & 4.74      & 31.72       & 6.50        &  & 4.39   & 3.70    \\
GG-ETS       & 330.38 & 75.54 & 37.02  & 87.39   & 78.11       & 25.64 & 52.71 &  & 25.69     & 24.84       & 32.09       &  & 113.47 & 58.13   \\
CT-ETS       & 96.73  & \textBF{4.71}  & 86.07  & 4.74    & 12.98       & 6.34  & 5.00  &  & 4.76      & 24.53       & 6.53        &  & 4.32   & 3.77    \\
             &        &       &        &         &             &       &       &  &           &             &             &  &        &         \\
{$h=5$}          &        &       &        &         &             &       &       &  &           &             &             &  &        &         \\
RT           & \textBF{15.18}  & 19.37 & \textBF{24.93}  & 19.08   & \textBF{8.57}  & 16.83 & 12.49 &  & 26.21     & 28.58       & 10.65       &  & 11.39  & 12.17   \\
GG-AR        & 310.90 & 87.30 & 77.45  & 91.95   & 71.53       & 76.92 & 81.72 &  & 90.74     & 86.24       & 89.78       &  & 94.33  & 70.19   \\
CT-AR        & 65.27  & 6.81  & 68.75  & 8.36    & 15.58       & 8.06  & 6.88  &  & 7.64      & 33.81       & 7.07        &  & 6.26   & 5.35    \\
GG-VAR       & 215.00 & 70.13 & 37.27  & 85.89   & 80.07       & 34.09 & 44.23 &  & 23.99     & 25.64       & 30.44       &  & 126.86 & 61.85   \\
CT-VAR       & 68.15  & 6.74  & 83.36  & 6.98    & 13.88       & \textBF{6.61}& \textBF{6.66}  &  & \textBF{6.75}  & \textBF{23.65}       & \textBF{5.90}        &  & 6.39   & 5.30    \\
GG-ARIMA     & 100.00 & 90.22 & 76.71  & 94.69   & 71.84       & 84.74 & 81.01 &  & 92.90     & 90.25       & 90.58       &  & 94.81  & 70.28   \\
CT-ARIMA     & 170.58 & 7.13  & 143.45 & 7.99    & 18.85       & 7.27  & 7.11  &  & 7.24      & 31.04       & 7.39        &  & 6.42   & \textBF{5.02}    \\
GG-ETS       & 287.44 & 90.45 & 76.41  & 95.59   & 71.33       & 84.89 & 81.34 &  & 92.64     & 88.06       & 90.62       &  & 94.96  & 70.33   \\
CT-ETS       & 98.07  & \textBF{6.72}  & 85.45  & \textBF{6.85}  & 13.92       & 6.91  & 6.78  &  & 7.12      & 24.10       & 7.40        &  & \textBF{6.22} & 5.13    \\
             &        &       &        &         &             &       &       &  &           &             &             &  &        &         \\
{$h=30$}        &        &       &        &         &             &       &       &  &           &             &             &  &        &         \\
RT           & \textBF{15.18}  & 19.37 & \textBF{24.93} & 19.08   & \textBF{8.57}   & 16.83 & 12.49 &  & 26.21     & 28.58       & 10.65       &  & 11.39  & 12.17   \\
GG-AR        & 310.89 & 85.36 & 77.77  & 89.22   & 71.32       & 76.78 & 81.39 &  & 90.54     & 87.34       & 90.08       &  & 95.13  & 70.25   \\
CT-AR        & 64.88  & 15.52 & 67.93  & 17.38   & 16.54       & 10.61 & 12.08 &  & \textBF{13.45}     & 31.35       & \textBF{9.07}    &  & \textBF{8.73}   & \textBF{7.60}   \\
GG-VAR       & 202.47 & 61.13 & 45.55  & 84.14   & 70.97       & 41.76 & 42.36 &  & 34.12     & 41.92       & 34.92       &  & 131.67 & 63.10   \\
CT-VAR       & 89.31  & \textBF{12.93} & 110.02 & \textBF{11.90}   & 25.74   & 13.62 & 11.72 &  & 16.55     & 41.51       & 9.91        &  & 11.36  & 9.00    \\
GG-ARIMA     & 100.00 & 87.52 & 77.15  & 91.22   & 72.28       & 83.19 & 80.56 &  & 92.96     & 90.13       & 91.54       &  & 95.70  & 70.38   \\
CT-ARIMA     & 159.50 & 17.00 & 141.90 & 17.32   & 21.31       & 11.71 & 13.89 &  & 14.03     & 27.41       & 12.21       &  & 9.83   & 7.90    \\
GG-ETS       & 287.44 & 87.72 & 76.80  & 92.16   & 71.29       & 83.49 & 80.90 &  & 92.72     & 89.59       & 91.50       &  & 95.96  & 70.39   \\
CT-ETS       & 105.94 & 14.05 & 92.63  & 14.56   & 16.60       & \textBF{9.59}  & \textBF{10.22} &  & 14.02 & \textBF{23.15}       & 10.76       &  & 10.16  & 8.14   
               \tabularnewline
\bottomrule
\end{tabular}
\par\end{centering}{\footnotesize \par}
\centering{}{\footnotesize This table presents the Mean Absolute Percentage Error (MAPE) metric across all moneyness and term structure data points available for the 12 commodity options in our sample during the 500-day out-of-sample period (December 2014-December 2016). Results are presented for 1-, 5-, and 30-day-ahead forecasts. The lowest forecast errors for each commodity and forecast horizon are shown in bold.}{\footnotesize \par}
\end{table}
{\tiny \par}

\begin{table}[!htbp]
\tabcolsep 0.03in
{\caption{Out-of-Sample SSR}\label{tab:OOS_MCPDC}
}{\footnotesize \par}
\begin{centering}
\begin{tabular}{@{}lcccccccccccccc@{}}
\toprule
{\footnotesize Forecasting} & \multicolumn{7}{c}{{Agricultural}} &  & \multicolumn{3}{c}{{Energy}} &  & \multicolumn{2}{c}{{Metals}}\tabularnewline
\cline{2-8} \cline{10-12} \cline{14-15} 
{\footnotesize Horizon (h)} & {\footnotesize Cocoa} & {\footnotesize Corn} & {\footnotesize Cotton} & {\footnotesize Soybean} & {\footnotesize Soybean} & {\footnotesize Sugar} & {\footnotesize Wheat} &  & {\footnotesize Crude} & {\footnotesize Heating} & {\footnotesize Natural} &  & {\footnotesize Gold} & {\footnotesize Silver}\tabularnewline
& & & & & {\footnotesize Oil} & & & &{\footnotesize Oil} & {\footnotesize Oil}  & {\footnotesize Gas} \\
\midrule
{$h=1$}      &        &        &        &         &             &       &       &  &           &             &             &  &        &         \\
RT       & \textBF{55\%}                 & 54\%                 & 53\%                 & 52\%                 & 56\%                 & 52\%                 & 58\%                 &  & 52\%                 & 51\%                 & 53\%                 &  & 59\%                 & 57\%                  \\
GG-AR    & 51\%                 & 54\%                 & 54\%                 & 53\%                 & 51\%                 & 51\%                 & 56\%                 &  & 59\%                 & 55\%                 & 58\%                 &  & 50\%                 & 52\%                  \\
CT-AR    & 52\%                 & 94\%                 & 52\%                 & 84\%                 & 56\%                 & 59\%                 & 88\%                 &  & 69\%                 & 51\%                 & 64\%                 &  & 88\%                 & 74\%                  \\
GG-VAR   & 52\%                 & 54\%                 & \textBF{55\%}                 & 53\%                 & 51\%                 & 53\%                 & 55\%                 &  & 63\%                 & 59\%                 & 57\%                 &  & 50\%                 & 52\%                  \\
CT-VAR   & 54\%                 & \textBF{95\%}                 & 51\%                 & \textBF{91\%}                 & \textBF{76\%}             & \textBF{78\%}               & \textBF{92\%}               &  & \textBF{94\%}                 & \textBF{61\%}                 & \textBF{85\%}           &  & \textBF{95\%}                 & \textBF{87\%}                  \\
GG-ARIMA & 50\%                 & 54\%                 & \textBF{55\%}                 & 53\%                 & 51\%                 & 55\%                 & 55\%                 &  & 61\%                 & 56\%                 & 58\%                 &  & 51\%                 & 52\%                  \\
CT-ARIMA & 53\%                 & 93\%                 & 51\%                 & 87\%                 & 60\%                 & 65\%                 & 85\%                 &  & 81\%                 & 52\%                 & 71\%                 &  & 89\%                 & 77\%                  \\
GG-ETS   & 50\%                 & 54\%                 & \textBF{55\%}                 & 53\%                 & 51\%                 & 55\%                 & 55\%                 &  & 61\%                 & 56\%                 & 58\%                 &  & 51\%                 & 52\%                  \\
CT-ETS   & 52\%                 & 94\%                 & 52\%                 & 90\%                 & 61\%                 & 63\%                 & 85\%                 &  & 81\%                 & 52\%                 & 71\%                 &  & 91\%                 & 77\%                  \\
         & \multicolumn{1}{l}{} & \multicolumn{1}{l}{} & \multicolumn{1}{l}{} & \multicolumn{1}{l}{} & \multicolumn{1}{l}{} & \multicolumn{1}{l}{} & \multicolumn{1}{l}{} &  & \multicolumn{1}{l}{} & \multicolumn{1}{l}{} & \multicolumn{1}{l}{} &  & \multicolumn{1}{l}{} & \multicolumn{1}{l}{}  \\
{$h=30$}     & \multicolumn{1}{l}{} & \multicolumn{1}{l}{} & \multicolumn{1}{l}{} & \multicolumn{1}{l}{} & \multicolumn{1}{l}{} & \multicolumn{1}{l}{} & \multicolumn{1}{l}{} &  & \multicolumn{1}{l}{} & \multicolumn{1}{l}{} & \multicolumn{1}{l}{} &  & \multicolumn{1}{l}{} & \multicolumn{1}{l}{}  \\
RT       & 57\%                 & 63\%                 & 56\%                 & 67\%                 & 70\%                 & 56\%                 & 70\%                 &  & 44\%                 & 49\%                 & 57\%                 &  & 73\%                 & 65\%                  \\
GG-AR    & 49\%                 & 51\%                 & 50\%                 & 50\%                 & 54\%                 & 62\%                 & 49\%                 &  & 43\%                 & 50\%                 & 46\%                 &  & 48\%                 & 47\%                  \\
CT-AR    & 49\%                 & 75\%                 & 56\%                 & 71\%                 & 71\%                 & 74\%                 & 80\%                 &  & 64\%                 & 51\%                 & 75\%                 &  & 86\%                 & 84\%                  \\
GG-VAR   & 58\%                 & 57\%                 & \textBF{59\%}                 & 59\%                 & 52\%                 & 56\%                 & 60\%                 &  & 72\%                 & 69\%                 & 69\%                 &  & 59\%                 & 58\%                  \\
CT-VAR   & \textBF{60\%}                 & 69\%                 & 54\%                 & 74\%                 & \textBF{78\%}                 & 78\%                 & 76\%                 &  & 81\%                 & \textBF{73\%}            & 70\%                 &  & 82\%                 & 83\%                  \\
GG-ARIMA & 53\%                 & 50\%                 & 51\%                 & 49\%                 & 55\%                 & 61\%                 & 49\%                 &  & 42\%                 & 49\%                 & 46\%                 &  & 47\%                 & 47\%                  \\
CT-ARIMA & 55\%                 & 92\%                 & 50\%                 & 82\%                 & 74\%                 & 78\%                 & 83\%                 &  & 89\%                 & 53\%                 & 84\%                 &  & 89\%                 & 87\%                  \\
GG-ETS   & 50\%                 & 50\%                 & 51\%                 & 49\%                 & 55\%                 & 61\%                 & 49\%                 &  & 42\%                 & 50\%                 & 46\%                 &  & 47\%                 & 47\%                  \\
CT-ETS   & 56\%                 & \textBF{98\%}                 & 57\%                 & \textBF{96\%}                 & 76\%           & \textBF{81\%}                 & \textBF{94\%}                 &  & \textBF{94\%}                 & 54\%                 & \textBF{86\%}            &  & \textBF{96\%}                 & \textBF{91\%}                  
\tabularnewline      
\bottomrule
\end{tabular}
\par\end{centering}{\tiny \par}
\centering{}{\footnotesize This table presents the Sign Success Ratio (SSR) metric across all moneyness and term structure data points available for the 12 commodity options in our sample during the 500-day out-of-sample period (December 2014-December 2016). The SSR metric shows us what percentage of the time each model correctly predicts the direction of implied volatility change. Results are presented for 1-, and 30-day-ahead forecasts. All models are benchmarked versus the \cite{chalamandaris2011important} fitted implied volatility surface from the previous period. What percentage of the time does it correctly predict the direction of implied volatility. The highest SSR for each commodity and forecast horizon are shown in bold.}{\footnotesize \par}
\end{table}
{\footnotesize \par}

\begin{table}[!hbtp]
{\caption{RMSE ratio by Moneyness and Maturity for CT-RW and GG-RW}\label{tab:RMSE_Ratio}
}{\footnotesize \par}
\begin{centering}
\tabcolsep 0.04in
\begin{tabular}{@{}lccccccccccccc@{}}
\toprule
& \multicolumn{6}{c}{\footnotesize Moneynesss} & & \multicolumn{6}{c}{\footnotesize Maturity} \\
\cline{2-7} \cline{9-14}
 & \multicolumn{2}{c}{\footnotesize OTM} & \multicolumn{2}{c}{\footnotesize ATM} & \multicolumn{2}{c}{\footnotesize ITM} & & \multicolumn{2}{c}{\footnotesize Short Term} & \multicolumn{2}{c}{\footnotesize Medium Term} & \multicolumn{2}{c}{\footnotesize Long Term} \tabularnewline
 & {\footnotesize CT-RW} & {\footnotesize GG-RW} & {\footnotesize CT-RW} & {\footnotesize GG-RW} & {\footnotesize CT-RW} & {\footnotesize GG-RW}  & & {\footnotesize CT-RW} & {\footnotesize GG-RW} & {\footnotesize CT-RW} & {\footnotesize GG-RW} & {\footnotesize CT-RW} & {\footnotesize GG-RW} \tabularnewline
\midrule
{$h=1$}      & \multicolumn{1}{l}{} & \multicolumn{1}{l}{} & \multicolumn{1}{l}{} & \multicolumn{1}{l}{} & \multicolumn{1}{l}{} & \multicolumn{1}{l}{} &  & \multicolumn{1}{l}{} & \multicolumn{1}{l}{} & \multicolumn{1}{l}{} & \multicolumn{1}{l}{} & \multicolumn{1}{l}{} & \multicolumn{1}{l}{}  \\
RT       &\textBF{0.19}& \textBF{0.25}  & 0.10  & 0.46  & 0.11  & 0.06  &  & 0.13  & 0.09  & 0.05  & 0.15  & 2.37  & 0.07  \\
GG-AR    & 0.46  & 0.58  & 0.16  & 0.73  & 0.57  & 0.30  &  & 0.28  & 0.21  & 0.22  & 0.63  & 18.00 & 0.56  \\
CT-AR    & 0.24  & 0.30  & \textBF{0.07}  & \textBF{0.33}  & \textBF{0.07}  & \textBF{0.04}  &  & \textBF{0.11}  & \textBF{0.08}  & \textBF{0.04}  & \textBF{0.11}  & 1.16  & 0.04  \\
GG-VAR   & 0.44  & 0.56  & 0.15  & 0.67  & 0.57  & 0.30  &  & 0.27  & 0.20  & 0.22  & 0.63  & 17.88 & 0.55  \\
CT-VAR   & 0.25  & 0.32  & 0.08  & 0.37  & 0.08  & \textBF{0.04}  &  & 0.12  & 0.09  & \textBF{0.04}  & 0.13  & \textBF{1.12}  & \textBF{0.03}  \\
GG-ARIMA & 0.44  & 0.56  & 0.11  & 0.50  & 0.54  & 0.29  &  & 0.22  & 0.16  & 0.21  & 0.59  & 17.96 & 0.56  \\
CT-ARIMA & 0.23  & 0.30  & 0.12  & 0.56  & 0.14  & 0.07  &  & 0.16  & 0.12  & 0.08  & 0.22  & 2.35  & 0.07  \\
GG-ETS   & 0.45  & 0.58  & 0.17  & 0.76  & 0.57  & 0.30  &  & 0.28  & 0.21  & 0.23  & 0.64  & 18.07 & 0.56  \\
CT-ETS   & 0.22  & 0.28  & 0.09  & 0.41  & 0.10  & 0.05  &  & 0.13  & 0.10  & 0.05  & 0.15  & 1.32  & 0.04  \\
         & \multicolumn{1}{l}{} & \multicolumn{1}{l}{} & \multicolumn{1}{l}{} & \multicolumn{1}{l}{} & \multicolumn{1}{l}{} & \multicolumn{1}{l}{} &  & \multicolumn{1}{l}{} & \multicolumn{1}{l}{} & \multicolumn{1}{l}{} & \multicolumn{1}{l}{} & \multicolumn{1}{l}{} & \multicolumn{1}{l}{}  \\
{$h=30$}     & \multicolumn{1}{l}{} & \multicolumn{1}{l}{} & \multicolumn{1}{l}{} & \multicolumn{1}{l}{} & \multicolumn{1}{l}{} & \multicolumn{1}{l}{} &  & \multicolumn{1}{l}{} & \multicolumn{1}{l}{} & \multicolumn{1}{l}{} & \multicolumn{1}{l}{} & \multicolumn{1}{l}{} & \multicolumn{1}{l}{}  \\
RT       & \textBF{0.02}  & \textBF{0.23}  & \textBF{0.02}  & 0.18  & 0.03  & 0.11  &  & \textBF{0.02}  & \textBF{0.14}  & 0.55  & 0.12  & 1.47  & 0.09  \\
GG-AR    & 0.09  & 1.01  & 0.10  & 0.72  & 0.13  & 0.51  &  & 0.08  & 0.51  & 2.42  & 0.53  & 9.85  & 0.58  \\
CT-AR    & 0.03  & 0.32  & \textBF{0.02}  & \textBF{0.16}  & \textBF{0.02}  & \textBF{0.09}  &  & \textBF{0.02}  & \textBF{0.14}  & \textBF{0.49}  & \textBF{0.11}  & \textBF{1.02}  & \textBF{0.06}  \\
GG-VAR   & 0.05  & 0.58  & 0.04  & 0.27  & 0.14  & 0.56  &  & 0.05  & 0.28  & 2.23  & 0.49  & 10.91 & 0.65  \\
CT-VAR   & 0.04  & 0.42  & 0.03  & 0.23  & 0.03  & 0.13  &  & 0.03  & 0.19  & 0.69  & 0.15  & 1.44  & 0.09  \\
GG-ARIMA & 0.09  & 1.01  & 0.09  & 0.70  & 0.13  & 0.50  &  & 0.08  & 0.48  & 2.35  & 0.52  & 10.11 & 0.60  \\
CT-ARIMA & 0.03  & 0.33  & 0.03  & 0.24  & 0.04  & 0.15  &  & 0.03  & 0.19  & 0.83  & 0.18  & 1.69  & 0.10  \\
GG-ETS   & 0.09  & 1.00  & 0.10  & 0.71  & 0.13  & 0.53  &  & 0.08  & 0.51  & 2.47  & 0.54  & 10.15 & 0.60  \\
CT-ETS   & 0.03  & 0.31  & 0.03  & 0.20  & 0.03  & 0.13  &  & 0.03  & 0.17  & 0.63  & 0.14  & 1.39  & 0.08                 
\tabularnewline
\bottomrule
\end{tabular}
\par\end{centering}{\footnotesize \par}
\centering{}{\scriptsize This table presents the Root Mean Squared Error (RMSE) ratio of each of the time-series models listed to the benchmark \cite{chalamandaris2011important} and \cite{goncalves2006predictable} frameworks with no change random walk parameters specified, CT-RW and GG-RW, respectively. The ratios are averaged across all 12 commodity contracts (where applicable), and broken down by level of moneyness and maturity. For table brevity, results are presented for 1- and 30-day-ahead forecasts only, split into OTM ($<=$95\%), ATM (97.5\% to 102.5\%), ITM ($>=$105\%), Short- (1 to 3 months), Medium- (3 to 6 months) and Long-term ($>=6$ months) contracts. The tests are conducted during the 500-day out-of-sample period (December 2014-December 2016) with results presented for 1- and 30-day ahead forecasts. Please note that there is not an equal number of implied volatility observations in each group. A value of less than 1 indicates that the respective model performs better than the CT-RW or GG-RW benchmark, and a value of greater than 1 indicates that the model performs worse than the CT-RW or GG-RW benchmark, based on the RMSE metric. The best performing model broken down by moneyness, maturity, forecast horizon and benchmark, is shown in bold.}{\footnotesize \par}
\end{table}

Table~\ref{tab:OOS_MCPDC} presents the SSR metric, which is the percentage of times the model correctly predicts the direction of the forecast. The metric is given across all moneyness and maturity observations for each commodity. The directional accuracy measure has an intuitive 50\% random chance benchmark. It does not necessarily translate into superior magnitude metrics but could be used to help determine signals for trading strategies . Each model is benchmarked versus the no change CT-RW, with the under performing GG-RW benchmark excluded for brevity. Overall, we find that in the majority of instances, the time-series models directionally outperform the random walk benchmark. Time-series modeling of parameters leads to directional advantages with a large number of instances of CT models achieving over 80\% SSRs. It is again observed that GG models do not perform as well as RT and CT, with SSRs as low as 42\% observed when forecasting crude oil options using a GG-ETS at a 30-day-ahead horizon. We also see in general that it is more difficult to predict the direction of change at longer horizons. There are 22 instances of SSRs falling below the intuitive 50\% benchmark at the 30-day-ahead forecasting horizon, including one CT model; the CT-AR forecast for Cocoa options with its 49\% SSR. We can also see distinction in terms of the time-series model specified, with the multivariate CT-VAR working well at the one-day-ahead forecasting horizon but the univariate CT-ETS showing improved directional forecasts at the 30-day-ahead horizon.

To further identify the source of the forecasting performance of the various models we break the results down into moneyness and maturity segments of the IVS. First, we define the OTM options group as those with moneyness levels of less than or equal to 95\%, ATM options group as those with moneyness between 97.5\% and 102.5\%, and ITM options group as those with moneyness of greater than or equal to 105\%. Second, for maturities, we define short-term as between one and three months, medium-term as between three months and six months, and long-term (where applicable) as greater than or equal to 6 months. Using these moneyness and expiration splits we calculate errors for each group and present the results in Table~\ref{tab:RMSE_Ratio}. The errors are expressed as the ratio of the RMSE for each time-series model to the RMSE of the benchmark random walk models aggregated across all 12 commodity options in our sample. A value of less than 1 indicates that a model outperforms the CT-RW or GG-RW benchmark, with a value of greater than 1 indicating that the model underperforms the benchmark. Overall, we see that the majority of the ratios are less than 1; indicating that the models outperform the benchmarks and that time-series modeling of the coefficients is effective.\footnote{The groups presented in Table 8 are very diverse in terms of their composition. This complicates a direct comparison of the ratio levels in a cross group evaluation. For instance, the groups range from short-term, that include 1 month maturity options, to long-term, that include 2 year maturity options, and are comprised of differing numbers of commodities and IV observations. For example, the short-term group has nine times more observations than the OTM group and the long-term group does not include four commodities from its calculation (as outlined in Table \ref{tab:Data-Description}).} We also see that the improvement over the benchmark is generally more pronounced at the longer 30-day-ahead forecasting horizon, with ratios as low as 0.02.

When we group the results by moneyness it is difficult to see any strong dynamics at play with, for example, the RT model showing relatively similar performance across all moneyness levels. More illuminating, however, is the maturity split. We find that while the time-series models perform well at predicting short- and medium- term maturity options, the CT-RW benchmark shows greater forecasting accuracy for long-term maturity options, indicated by all RMSE ratios being greater than 1. This highlights the difficulty of beating the random walk benchmark across the entire IVS and also indicates that time-series modeling of parameters is not as relevant for long-term maturity models as the explicit modeling of the term structure component.

\begin{table}[!htbp]
\tabcolsep 0.043in
{\caption{Model Confidence Set Results}\label{tab:MCS_Results}
}{\footnotesize \par}
\begin{centering}
\begin{tabular}{@{}lcccccccccccccc@{}}
\toprule
{Test Statistic} & \multicolumn{7}{c}{{Agricultural}} &  & \multicolumn{3}{c}{{Energy}} &  & \multicolumn{2}{c}{{Metals}}\tabularnewline
\cline{2-8} \cline{10-12} \cline{14-15} 
{$T_{\text{R,M}}$} & {\footnotesize Cocoa} & {\footnotesize Corn} & {\footnotesize Cotton} & {\footnotesize Soybean} & {\footnotesize Soybean} & {\footnotesize Sugar} & {\footnotesize Wheat} &  & {\footnotesize Crude} & {\footnotesize Heating} & {\footnotesize Natural} &  & {\footnotesize Gold} & {\footnotesize Silver}\tabularnewline
& & & & & {\footnotesize Oil} & & &  & {\footnotesize Oil} & {\footnotesize Oil} & {\footnotesize Gas} \\
\midrule
{$h=1$} &  &  &  &  &  &  &  &  &  &  &   &  &  & \tabularnewline
RT       & $\dagger$ &   & $\dagger$ &   & $\dagger$ &   &   &   &   &   &   &   &   &    \\
CT-AR    &   & $\dagger$ &   &   &   &   & $\dagger$ &   &   &   &   &   &   &    \\
CT-VAR   &   & $\dagger$ &   & $\dagger$ &   &   &   &   & $\dagger$ &   & $\dagger$ &   & $\dagger$ & $\dagger$  \\
CT-ETS   &   & $\dagger$ &   &   &   & $\dagger$ &   &   &   &   &   &   & $\dagger$ & $\dagger$  \\
CT-ARIMA &   &   &   &   &   & $\dagger$ &   &   &   &   &   &   &   & $\dagger$  \\
CT-RW    &   &   &   & $\dagger$ & $\dagger$ & $\dagger$ & $\dagger$ &   &   &   &   &   &   & $\dagger$  \\
GG-AR    &   &   &   &   &   &   &   &   &   &   &   &   &   &    \\
GG-VAR   &   &   &   &   &   &   &   &   &   &   &   &   &   &    \\
GG-ETS   &   &   &   &   &   &   &   &   &   &   &   &   &   &    \\
GG-ARIMA &   &   &   &   &   &   &   &   &   & $\dagger$ &   &   &   &    \\
GG-RW    &   &   &   &   &   &   &   &   &   &   &   &   &   &                      \\
                   &            &            &            &            &            &             &            &  &             &            &            &                      &                      &                       \\
                   {$h=5$}      &   &   &   &   &   &   &   &  &   &   &   &  &   &    \\
RT       & $\dagger$ &   & $\dagger$ &   & $\dagger$ &   &   &  &   & $\dagger$ &   &  &   &    \\
CT-AR    &   & $\dagger$ &   &   &   &   &   &  &   &   &   &  & $\dagger$ &    \\
CT-VAR   &   & $\dagger$ &   & $\dagger$ &   &   & $\dagger$ &  & $\dagger$ &   & $\dagger$ &  & $\dagger$ &    \\
CT-ETS   &   &   &   &   &   & $\dagger$ &   &  &   &   &   &  & $\dagger$ &    \\
CT-ARIMA &   &   &   &   &   & $\dagger$ &   &  &   &   &   &  &   & $\dagger$  \\
CT-RW    &   &   &   &   &   & $\dagger$ &   &  & $\dagger$ &   &   &  &   &    \\
GG-AR    &   &   &   &   &   &   &   &  &   &   &   &  &   &    \\
GG-VAR   &   &   &   &   &   &   &   &  &   &   &   &  &   &    \\
GG-ETS   &   &   &   &   &   &   &   &  &   &   &   &  &   &    \\
GG-ARIMA &   &   &   &   &   &   &   &  &   &   &   &  &   &    \\
GG-RW    &   &   &   &   &   &   &   &  &   &   &   &  &   & \\
                   &            &            &            &            &            &             &            &  &             &            &            &                      &                      &                       \\

{$h=30$}     &   &   &   &   &   &   &   &   &   &   &   &   &   &    \\
RT       & $\dagger$ &   & $\dagger$ &   & $\dagger$ &   &   &   &   & $\dagger$ &   &   &   &    \\
CT-AR    &   &   &   &   &   &   &   &   & $\dagger$ &   & $\dagger$ &   & $\dagger$ & $\dagger$  \\
CT-VAR   &   & $\dagger$ &   & $\dagger$ &   &   &   &   &   &   &   &   &   &    \\
CT-ETS   &   &   &   &   &   & $\dagger$ & $\dagger$ &   & $\dagger$ & $\dagger$ &   &   &   &    \\
CT-ARIMA &   &   &   &   &   &   &   &   & $\dagger$ &   &   &   &   &    \\
CT-RW    &   &   &   &   &   & $\dagger$ &   &   & $\dagger$ & $\dagger$ &   &   &   &   \\
GG-AR    &   &   &   &   &   &   &   &   &   &   &   &   &   &    \\
GG-VAR   &   &   &   &   &   &   &   &   &   &   &   &   &   &    \\
GG-ETS   &   &   &   &   &   &   &   &   &   &   &   &   &   &    \\
GG-ARIMA &   &   &   &   &   &   &   &   &   &   &   &   &   &                        \\
GG-RW    &   &   &   &   &   &   &   &   &   &   $\dagger$ &   &   &   &                      \\
\bottomrule
\end{tabular}
\par\end{centering}{\footnotesize \par}
\centering{}{\footnotesize 
This table presents the results of the model confidence set test of \cite{HLN11} based on the out-of-sample forecasts produced by various models across the full implied volatility surface for each of the 12 commodity options in our sample. It is assessed during the 500-day out-of-sample period (December 2014-December 2016), with results being presented for 1-, 5- and 30-day-ahead forecasts. The $\dagger$ symbol is used to indicate that a particular model resides in the superior set of models for a given commodity and forecast horizon.}{\footnotesize \par}
\end{table}

Having focused on a direct comparison of forecasting errors to date, we now formally investigate cross-model superiority using the MCS of \cite{HLN11}. The goal is to identify what model(s) reside in the superior set of models to establish if any statistically outperform the others. It also addresses the issue of possible false discoveries arising from the multiple comparisons problem. Overall, the results in Table \ref{tab:MCS_Results} indicate that all three frameworks perform well in terms of identifying predictability, with even the GG-ARIMA residing in the superior set of models for one-day-ahead forecasts of Heating Oil. 

Despite no random walk benchmark solely residing in the superior set of models for any commodity or forecasting horizon, the simultaneous appearance of the benchmarks alongside other models highlights how difficult they are to beat. More specifically, for the short one-day-ahead forecasting horizon the random walk benchmarks are statistically indistinguishable from the time-series process specified models across five commodities, Soybean, Soybean Oil, Sugar, Wheat and Silver. However, for longer forecasting horizons the strength of this dynamic subsides, with the benchmarks only appearing in the superior set of models for two commodities at the 5-day ahead horizon. This indicates that it takes a significant passing of time to see a formal benefit from the second step of our model construction in which we explicitly specify parameter evolution.

When we turn our attention to the commodity group split it can be seen that for Precious Metals the CT models are the only models to exhibit formal outperformance. Conversely, the RT model performs best when focusing on Agricultural options, despite results here being more mixed, with the RT model appearing in the superior set of models for Cocoa, Cotton, and Soybean Oil across all horizons. The CT models reside in the superior set of models for Wheat and Soybean, commodities associated with extreme average convenience yield slopes, indicating that those commodities with steeper convenience yield curves are more accurately forecasted using the explicit Nelson-Siegel CT framework.

\section{Conclusion}\label{sec:Conclusions}

Inter-temporal predictability in the IVSs of equity, index, foreign exchange, and interest rate options has been uncovered using a small number of models. Motivated by the growing popularity of commodities and the increasing correlation between commodities and other markets, this paper employs these existing frameworks to ascertain if they can both characterize and forecast the IVSs observed for popular commodity options. IV smile asymmetry is observed across commodities with fear relating to a supply-side shock being the most prevalent characteristic for agricultural options. In a cross-model comparison of the multiple parametric factor and machine learning models considered, no model systematically outperforms its competitors. For this reason, we draw inferences from what models demonstrate outperformance for specific groups of commodities.

Focusing on out-of-sample testing the most striking conclusion is that when modeling Energy and Precious Metals options the CT specification produces the most accurate forecasts. This aligns with \cite{doran2008computing} who highlight that there is a significant term structure of volatility present in Energy markets even at short-term maturities. The finding is at odds with what \cite{chalamandaris2011important} conclude for foreign exchange options; that a linear approximation of the term structure is sufficient up to an option expiry of 12 months ahead. Precious Metals, however, is a class of commodities that exhibits relatively low levels of IV for longer maturity dates, in contrast to what is observed for other commodities. We hypothesize that this converse maturity dynamic is caused by safe haven properties associated with Precious Metals; properties not commonly attributed to Agricultural or Energy commodities.\footnote{\cite{baur2010gold} empirically demonstrate the short-term safe haven attribute associated with gold.} Again, the explicit modeling of the term structure using the Nelson-Siegel factors leads to CT exhibiting improved predictability for long-term maturities. We also find that cross-model outperformance is less stark for Agricultural options. However, the RT model exhibits the most promising results, in particular for commodities with flatter underlying convenience yield curves. Further, we conclude that the simpler specification of the univariate models over the multivariate VAR model leads, on average to, superior forecasting results at longer forecasting horizons, as also established for foreign exchange options in \cite{Chalamandaris2014}.

The mapping between the approaches we consider, and the optimizing behavior of a market participant who experiences incomplete information in a rational asset pricing framework, is complex. Despite this, the existence of a precisely estimable dynamic relationship could still be explained by general equilibrium models, whereby a learning process followed by investors leads to observed inter-temporal IVS dependence, as hypothesized by \cite{bernales2015learning}. Additionally, despite not being the focus of this performance paper, the pockets of predictability uncovered could be exploited economically in future work. However, this would involve the implementation of bespoke hedging or trading strategies dependent on each market participant's specific option focus and portfolio requirements.\footnote{Another reason we refrain from presenting a trading strategy is the inability to realistically simulate a live market environment. Previous studies rely on idiosyncratic market assumptions, and ignore implementation issues such as liquidity, strategy drawdowns, margin calls, bid-ask spread considerations, and microstructure effects that might distort any calculated profits.} Finally, as a potential avenue for future research, a forecast combination approach along the lines of \cite{BG69} and \cite{gogolin2016does} may yield a further refinement of forecast accuracy using the superior approaches identified here.

\if0\blind
{
\section*{Acknowledgements}

We thank Andrianos Tsekrekos, Donal McKillop, Chardin Wese Simen, Andrew Vivian, Georgios Sermipinis, Georgios Panos, and Andreas Hoepner, for comments and advice that have helped to improve this paper. Thanks also go to the seminar participants at a departmental seminar at Queens's University Belfast, a Wards Finance Seminar at the University of Glasgow, and conference participants at the FMA European Conference 2017, British Accounting and Finance Association Annual Conference 2018, and Infiniti Finance Conference 2017. This final version has been greatly improved by the comments of the editor and two referees. Han Lin Shang\textquoteright s work was supported in part by a Research School Grant from the ANU College of Business and Economics.
} \fi

\newpage
\bibliographystyle{elsarticle-harv}
\bibliography{11Feb17}

\end{document}